\pdfoutput=1
\documentclass[12pt,a4paper]{article}
\usepackage{ifthen} 
\newboolean{pdflatex}
\setboolean{pdflatex}{true}
\newboolean{articletitles}
\setboolean{articletitles}{true} 
\newboolean{uprightparticles}
\setboolean{uprightparticles}{false} 
\newboolean{inbibliography}
\setboolean{inbibliography}{false} 

\def\paperauthors{M.~Andreotti, S.~Capelli, G.~Cavallero, S.~Chiozzi, A.~Cotta~Ramusino, C.~D'Ambrosio, M.~Fiorini, E.~Franzoso, C.~Frei, S.~Gallorini, S.~Gambetta, C.~Giugliano, C.~Gotti, T.~Gys, F.~Keizer, M.~Maino, B.~Malecki, L.~Minzoni, S.~Mitchell, I.~Neri, A.~Petrolini, D.~Piedigrossi, G.~Robertson, A.~Sergi, G.~Simi, I.~Slazyk, M.~Smith, J.~Webster, S.A.~Wotton}
\def\papertitle{Characterisation of signal-induced noise in Hamamatsu R11265 Multianode Photomultiplier Tubes} 
\def\paperasciititle{\papertitle} 
\def\paperkeywords{{Photon detectors for UV, visible and IR photons (vacuum) (photomultipliers, HPDs, others)}, {Cherenkov detectors}, {Performance of High Energy Physics Detectors},
  {Detector modelling and simulations II (electric fields, charge transport, multiplication and induction, pulse formation, electron emission, etc)}}
\def\papercopyright{\the\year\ CERN for the benefit of the LHCb collaboration} 
\def\paperlicence{CC BY 4.0 licence}
\def\paperlicenceurl{https://creativecommons.org/licenses/by/4.0/}

\usepackage[top=1in, bottom=1.25in, left=1in, right=1in]{geometry}

\usepackage{microtype}
\usepackage{lineno}  
\usepackage{xspace} 
\usepackage{caption} 

\usepackage{graphicx}  
\usepackage{color}
\usepackage{colortbl}
\graphicspath{{./figs/}} 

\usepackage{amsmath} 
\usepackage{amssymb}
\usepackage{amsfonts}
\usepackage{upgreek} 

\newcommand*\patchAmsMathEnvironmentForLineno[1]{%
\expandafter\let\csname old#1\expandafter\endcsname\csname #1\endcsname
\expandafter\let\csname oldend#1\expandafter\endcsname\csname
end#1\endcsname
 \renewenvironment{#1}%
   {\linenomath\csname old#1\endcsname}%
   {\csname oldend#1\endcsname\endlinenomath}%
}
\newcommand*\patchBothAmsMathEnvironmentsForLineno[1]{%
  \patchAmsMathEnvironmentForLineno{#1}%
  \patchAmsMathEnvironmentForLineno{#1*}%
}
\AtBeginDocument{%
\patchBothAmsMathEnvironmentsForLineno{equation}%
\patchBothAmsMathEnvironmentsForLineno{align}%
\patchBothAmsMathEnvironmentsForLineno{flalign}%
\patchBothAmsMathEnvironmentsForLineno{alignat}%
\patchBothAmsMathEnvironmentsForLineno{gather}%
\patchBothAmsMathEnvironmentsForLineno{multline}%
\patchBothAmsMathEnvironmentsForLineno{eqnarray}%
}

\usepackage{hyperxmp}

\usepackage[pdftex,
            pdfauthor={\paperauthors},
            pdftitle={\paperasciititle},
            pdfkeywords={\paperkeywords},
            pdfcopyright={Copyright (C) \papercopyright},
            pdflicenseurl={\paperlicenceurl}]{hyperref}

\usepackage[colorinlistoftodos,textsize=scriptsize]{todonotes}

\usepackage[bottom,flushmargin,hang,multiple]{footmisc}

\usepackage[all]{hypcap} 

\usepackage{xspace} 
\usepackage{upgreek}

\def\lhcb   {\mbox{LHCb}\xspace}

\def\cern {\mbox{CERN}\xspace}
\def\lhc    {\mbox{LHC}\xspace}

\def\rich   {RICH\xspace}
\def\richone {RICH1\xspace}
\def\richtwo {RICH2\xspace}

\def\MagUp {\mbox{\em Mag\kern -0.05em Up}\xspace}

\ifthenelse{\boolean{uprightparticles}}%
{

 \def\PDelta      {\ensuremath{\Delta}\xspace}                 
 \def\PXi         {\ensuremath{\Xi}\xspace}                 
 \def\PLambda     {\ensuremath{\Lambda}\xspace}                 
 \def\PSigma      {\ensuremath{\Sigma}\xspace}                 
 \def\POmega      {\ensuremath{\Omega}\xspace}                 
 \def\PUpsilon    {\ensuremath{\Upsilon}\xspace}

 \def\PB      {\ensuremath{\mathrm{B}}\xspace}                 
                  
 \def\PD      {\ensuremath{\mathrm{D}}\xspace}

 \def\PK      {\ensuremath{\mathrm{K}}\xspace}

 \def\Pi      {\ensuremath{\mathrm{i}}\xspace}

 \def\Ps      {\ensuremath{\mathrm{s}}\xspace}

 \def\thebaroffset{0.0em}
}
{

 \mathchardef\PDelta="7101
 \mathchardef\PXi="7104
 \mathchardef\PLambda="7103
 \mathchardef\PSigma="7106
 \mathchardef\POmega="710A
 \mathchardef\PUpsilon="7107
                  
 \def\PB      {\ensuremath{B}\xspace}                 
                  
 \def\PD      {\ensuremath{D}\xspace}

 \def\PK      {\ensuremath{K}\xspace}

 \def\Pi      {\ensuremath{i}\xspace}

 \def\Ps      {\ensuremath{s}\xspace}

 \def\thebaroffset{0.18em}
}
\newcommand{\offsetoverline}[2][\thebaroffset]{\kern #1\overline{\kern -#1 #2}}%

\makeatletter
\ifcase \@ptsize \relax
  \newcommand{\miniscule}{\@setfontsize\miniscule{4}{5}}
\or
  \newcommand{\miniscule}{\@setfontsize\miniscule{5}{6}}
\or
  \newcommand{\miniscule}{\@setfontsize\miniscule{5}{6}}
\fi
\makeatother

\DeclareRobustCommand{\optbar}[1]{\shortstack{{\miniscule (\rule[.5ex]{1.25em}{.18mm})}
  \\ [-.7ex] $#1$}}

\def\squark    {{\ensuremath{\Ps}}\xspace}

\def\KorKbar {\kern \thebaroffset\optbar{\kern -\thebaroffset \PK}{}\xspace}

\def\D       {{\ensuremath{\PD}}\xspace}

\def\DorDbar {\kern \thebaroffset\optbar{\kern -\thebaroffset \PD}\xspace}

\def\Dp      {{\ensuremath{\D^+}}\xspace}
\def\Dm      {{\ensuremath{\D^-}}\xspace}

\def\DpDm    {\ensuremath{\Dp {\kern -0.16em \Dm}}\xspace}

\def\B       {{\ensuremath{\PB}}\xspace}

\def\BorBbar {\kern \thebaroffset\optbar{\kern -\thebaroffset \PB}\xspace}

\def\Bd      {{\ensuremath{\B^0}}\xspace}

\def\BdorBdbar {\kern \thebaroffset\optbar{\kern -\thebaroffset \Bd}\xspace}

\def\Bs      {{\ensuremath{\B^0_\squark}}\xspace}

\def\BsorBsbar {\kern \thebaroffset\optbar{\kern -\thebaroffset \Bs}\xspace}

\def\Y#1S{\ensuremath{\PUpsilon{(#1S)}}\xspace}

\def\LorLbar     {\kern \thebaroffset\optbar{\kern -\thebaroffset \PLambda}\xspace}

\def\CP                {{\ensuremath{C\!P}}\xspace}

\def\AT#1     {\ensuremath{A_{\mathrm{T}}^{#1}}\xspace}           

\def\C#1      {\ensuremath{\mathcal{C}_{#1}}\xspace}                       
\def\Cp#1     {\ensuremath{\mathcal{C}_{#1}^{'}}\xspace}                    
\def\Ceff#1   {\ensuremath{\mathcal{C}_{#1}^{\mathrm{(eff)}}}\xspace}        
\def\Cpeff#1  {\ensuremath{\mathcal{C}_{#1}^{'\mathrm{(eff)}}}\xspace}       
\def\Ope#1    {\ensuremath{\mathcal{O}_{#1}}\xspace}                       
\def\Opep#1   {\ensuremath{\mathcal{O}_{#1}^{'}}\xspace}                    

\newcommand{\nospaceunit}[1]{\ensuremath{\text{#1}}}       
\newcommand{\aunit}[1]{\ensuremath{\text{\,#1}}}       
\newcommand{\unit}[1]{\aunit{#1}\xspace}                   

\newcommand{\tev}{\aunit{Te\kern -0.1em V}\xspace}
\newcommand{\gev}{\aunit{Ge\kern -0.1em V}\xspace}
\newcommand{\mev}{\aunit{Me\kern -0.1em V}\xspace}
\newcommand{\kev}{\aunit{ke\kern -0.1em V}\xspace}
\newcommand{\ev}{\aunit{e\kern -0.1em V}\xspace}
 
\newcommand{\mevc}{\ensuremath{\aunit{Me\kern -0.1em V\!/}c}\xspace}
\newcommand{\gevc}{\ensuremath{\aunit{Ge\kern -0.1em V\!/}c}\xspace}
\newcommand{\mevcc}{\ensuremath{\aunit{Me\kern -0.1em V\!/}c^2}\xspace}
\newcommand{\gevcc}{\ensuremath{\aunit{Ge\kern -0.1em V\!/}c^2}\xspace}

\def\cm   {\aunit{cm}\xspace}
\def\cma  {\ensuremath{\aunit{cm}^2}\xspace}
\def\mm   {\aunit{mm}\xspace}
\def\mma  {\ensuremath{\aunit{mm}^2}\xspace}

\def\sec  {\ensuremath{\aunit{s}}\xspace}

\def\mus  {\ensuremath{\,\upmu\nospaceunit{s}}\xspace}
\def\ns   {\ensuremath{\aunit{ns}}\xspace}

\def\mhz  {\ensuremath{\aunit{MHz}}\xspace}
\def\khz  {\ensuremath{\aunit{kHz}}\xspace}

\def\gsim{{~\raise.15em\hbox{$>$}\kern-.85em
          \lower.35em\hbox{$\sim$}~}\xspace}
\def\lsim{{~\raise.15em\hbox{$<$}\kern-.85em
          \lower.35em\hbox{$\sim$}~}\xspace}

\def\boole      {\mbox{\textsc{Boole}}\xspace}

\def\tell1  {TELL1\xspace}
\def\ukl1   {UKL1\xspace}

\def\cffour        {\ensuremath{\mathrm{ CF_4}}\xspace}

\newcommand{\eg}{\mbox{\itshape e.g.}\xspace}
\newcommand{\ie}{\mbox{\itshape i.e.}\xspace}

\newcommand{\etc}{\mbox{\itshape etc.}\xspace}

\usepackage{cite} 
\usepackage{mciteplus}

\usepackage{longtable} 
\usepackage{subfigure}
\usepackage{booktabs}
\usepackage{dcolumn}
\usepackage{comment}

\begin{document}

\renewcommand{\thefootnote}{\fnsymbol{footnote}}
\setcounter{footnote}{1}

\begin{titlepage}
\pagenumbering{roman}
\vspace*{-1.5cm}
\centerline{\large EUROPEAN ORGANIZATION FOR NUCLEAR RESEARCH (CERN)}
\vspace*{1.5cm}
\noindent
\begin{tabular*}{\linewidth}{lc@{\extracolsep{\fill}}r@{\extracolsep{0pt}}}
\ifthenelse{\boolean{pdflatex}}
{\vspace*{-1.5cm}\mbox{\!\!\!\includegraphics[width=.14\textwidth]{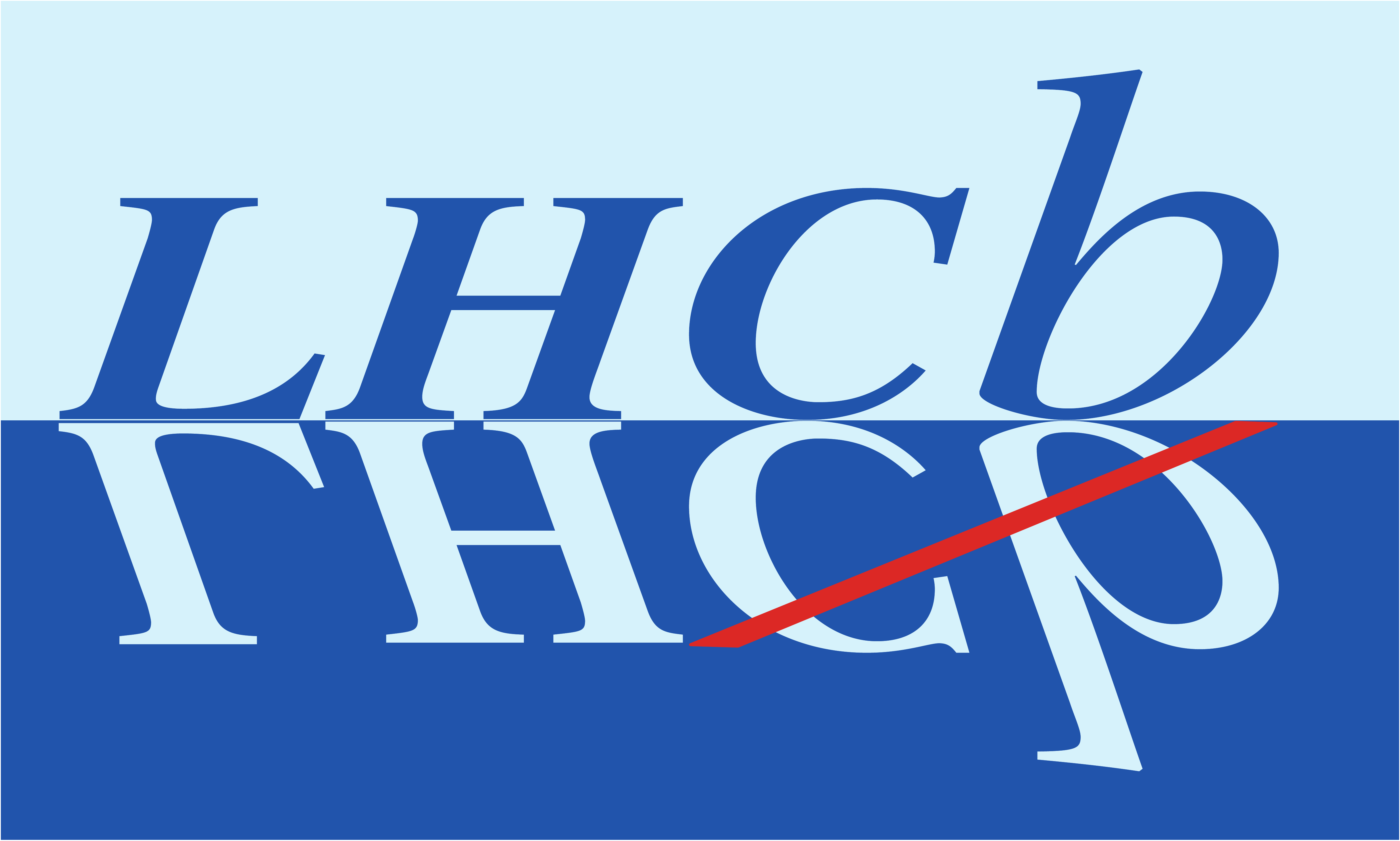}} & &}
{\vspace*{-1.2cm}\mbox{\!\!\!\includegraphics[width=.12\textwidth]{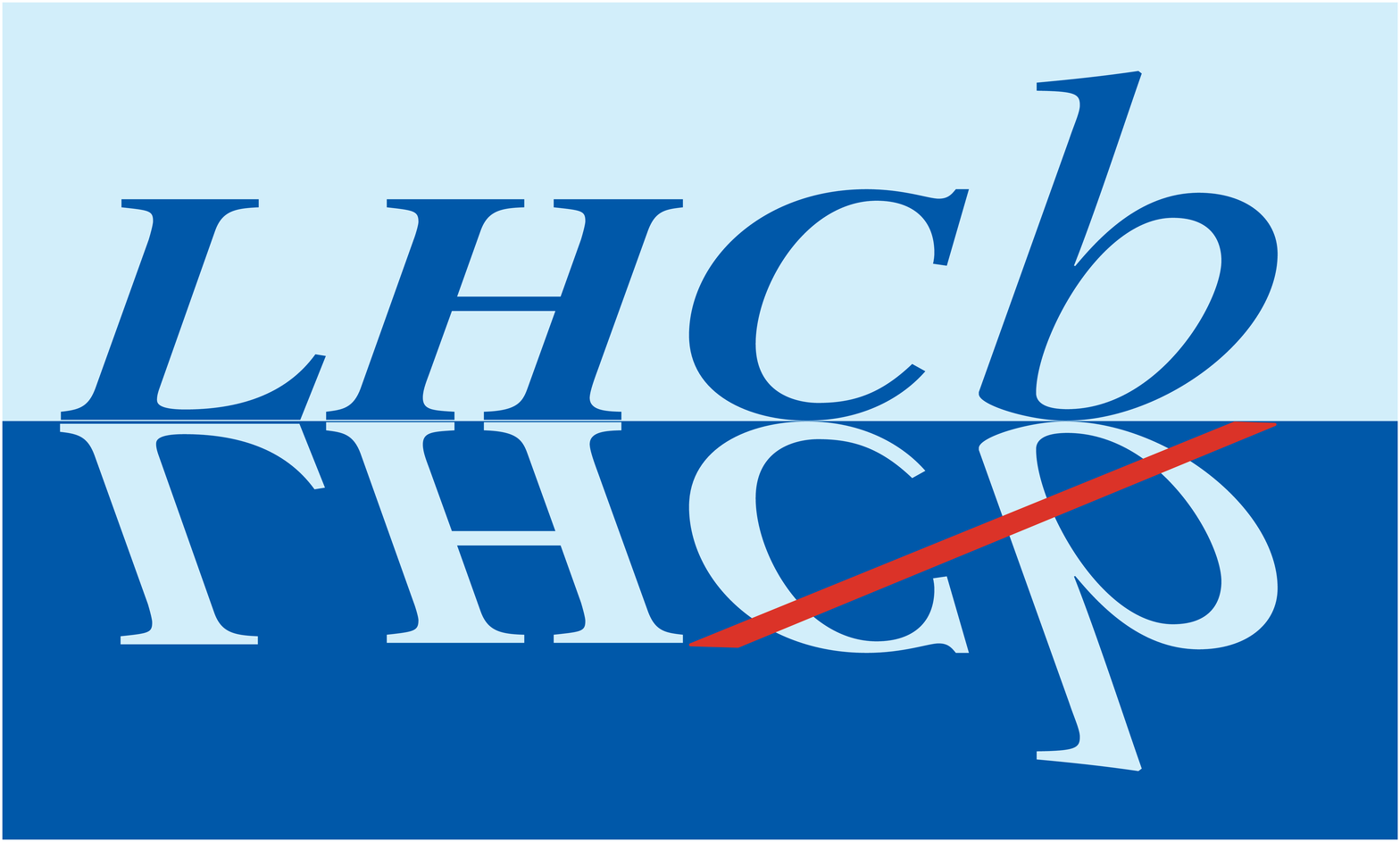}} & &} 
\\
 & & CERN-LHCb-DP-2021-005 \\ 
 & & November 23, 2021 \\ 
 & & \\
\end{tabular*}

\vspace*{1.0cm}

{\normalfont\bfseries\boldmath\huge
\begin{center}
  \papertitle 
\end{center}
}

\vspace*{0.5cm}

\begin{center}
\centerline
\small
M.~Andreotti$^{1}$,
S.~Capelli$^{2,a}$,
G.~Cavallero$^{3,*}$,
S.~Chiozzi$^{1}$,
A.~Cotta~Ramusino$^{1}$,
C.~D'Ambrosio$^{3}$,
M.~Fiorini$^{1,b}$,
E.~Franzoso$^{1,b}$,
C.~Frei$^{3}$,
S.~Gallorini$^{\dag}$,
S.~Gambetta$^{4,3}$,
C.~Giugliano$^{1,b}$,
C.~Gotti$^{2}$,
T.~Gys$^{3}$,
F.~Keizer$^{3}$,
M.~Maino$^{\dag\dag}$,
B.~Malecki$^{3}$,
L.~Minzoni$^{\dag\dag\dag}$,
S.~Mitchell$^{4}$,
I.~Neri$^{1}$,
A.~Petrolini$^{5,c}$,
D.~Piedigrossi$^{3}$,
G.~Robertson$^{4}$,
A.~Sergi$^{5,c}$,
G.~Simi$^{6}$,
I.~Slazyk$^{1,b}$,
M.~Smith$^{7}$,
J.~Webster$^{4}$,
S.A.~Wotton$^{8}$.\bigskip

{\footnotesize \it

$^{1}$INFN Sezione di Ferrara, Ferrara, Italy\\
$^{2}$INFN Sezione di Milano-Bicocca, Milano, Italy\\
$^{3}$European Organization for Nuclear Research (CERN), Geneva, Switzerland\\
$^{4}$School of Physics and Astronomy, University of Edinburgh, Edinburgh, United Kingdom\\
$^{5}$INFN Sezione di Genova, Genova, Italy\\
$^{6}$Universita degli Studi di Padova, Universita e INFN, Padova, Padova, Italy\\
$^{7}$Imperial College London, London, United Kingdom\\
$^{8}$Cavendish Laboratory, University of Cambridge, Cambridge, United Kingdom\\
$^{a}$Universit{\`a} di Milano Bicocca, Milano, Italy\\
$^{b}$Universit{\`a} di Ferrara, Ferrara, Italy\\
$^{c}$Universit{\`a} di Genova, Genova, Italy\\
$^\dag$ Formerly at $^{6}$\\
$^{\dag\dag}$ Formerly at $^{2,a}$\\
$^{\dag\dag\dag}$ Formerly at $^{1,b}$\\
\vspace{1mm}
$^*$Corresponding author\\
}
\end{center}

\vspace{\fill}

\begin{abstract}
  \noindent
  Signal-induced noise is observed in Hamamatsu R11265 Multianode Photomultiplier Tubes, manifesting up to several microseconds after the single photoelectron response signal and localised in specific anodes. The mean number of noise pulses varies between devices, and shows significant dependence on the applied high-voltage. The characterisation of this noise and the mitigation strategies to perform optimal single-photon counting at 40\mhz, as required by the \lhcb Ring-Imaging Cherenkov detectors, are reported.
\end{abstract}

\vspace{\fill}

\begin{center}
  Published in JINST 16 (2021) P11030
\end{center}

\vspace{\fill}

{\footnotesize 
\centerline{\copyright~\papercopyright. \href{\paperlicenceurl}{\paperlicence}.}}
\vspace*{2mm}

\end{titlepage}

\newpage
\setcounter{page}{2}
\mbox{~}

\renewcommand{\thefootnote}{\arabic{footnote}}
\setcounter{footnote}{0}

\cleardoublepage

\pagestyle{plain} 
\setcounter{page}{1}
\pagenumbering{arabic}

\section{Introduction}
\label{sec:Introduction}

The \lhcb experiment~\cite{LHCb-DP-2008-001} at the Large Hadron Collider (LHC) is designed to search for New Physics beyond the Standard Model in the study of \CP violating processes and rare $b$- and $c$-hadron decays. The excellent performance of the \lhcb detector during LHC Run 1 and 2 allowed to extend the core physics programme to include additional flavour-physics measurements, such as lepton flavour universality tests and conventional and exotic hadron spectroscopy~\cite{LHCb-DP-2014-002}.

Two Ring-Imaging Cherenkov (\rich) detectors~\cite{LHCb-DP-2012-003} provide long-lived charged hadron discrimination in the momentum range 2-100 \gevc. This particle identification (PID) capability allows to reduce the combinatorial background and to discriminate between decay modes of otherwise identical topology: the \rich detectors are therefore crucial to fulfil the flavour-physics programme of \lhcb~\cite{LHCb-DP-2012-003}.

The \lhcb experiment is undergoing a major upgrade~\cite{LHCb-TDR-012} to enable operation at an instantaneous luminosity of $2\times 10^{33} \cm^{-2} \sec^{-1}$ during the LHC Run 3 which will start in 2022. To fully exploit the flavour-physics potential with this five-fold increase in the instantaneous luminosity, the whole \lhcb detector will be read out at the full 40\mhz LHC bunch crossing rate and a flexible full software trigger~\cite{LHCb-TDR-016} will be implemented. The Hybrid Photon Detectors (HPDs) used in the \rich system during LHC Run 1 and 2, with embedded readout electronics working at 1\mhz trigger rate~\cite{LHCb-DP-2008-001}, will be replaced with Multianode Photomultiplier Tubes (MaPMTs) and new external Front-End (FE) electronics~\cite{LHCb-TDR-014}, able to provide single photon counting capability at the 40\mhz bunch crossing rate. The high-rate and high-occupancy environment imposes strict requirements on the properties of the photon detectors, including the demand for a low count rate caused by dark noise and other sources of internal instrumental noise. The opto-electronics chain that will be used in the upgraded \lhcb \rich detectors is described in Sec.~\ref{sec:mapmtDescription}. 

\begin{figure}[tb]
   \centering
\subfigure[]{
 \includegraphics[width=6cm]{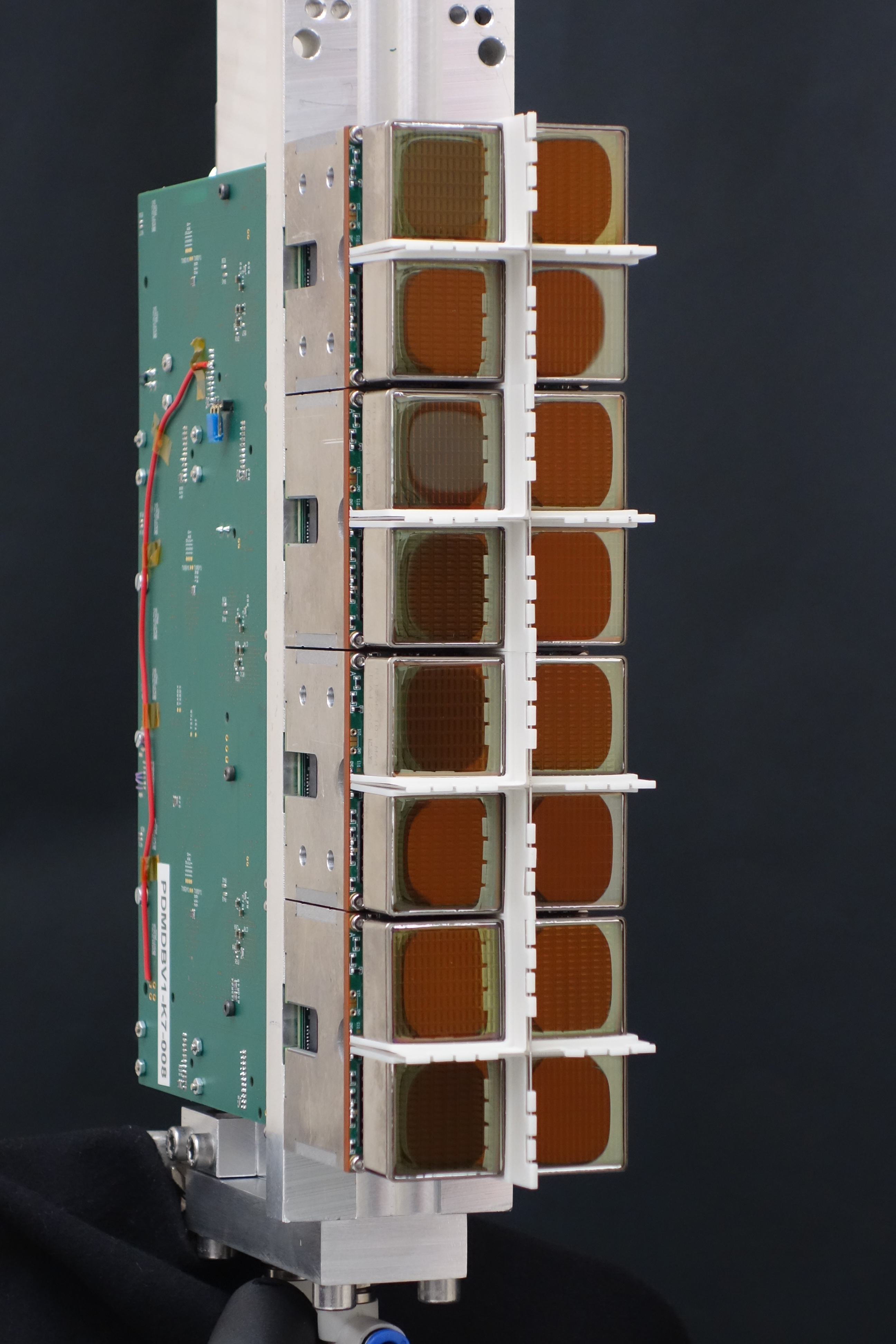}
     \label{PDMpicture}}
   \subfigure[]{
     \includegraphics[width=8cm, keepaspectratio]{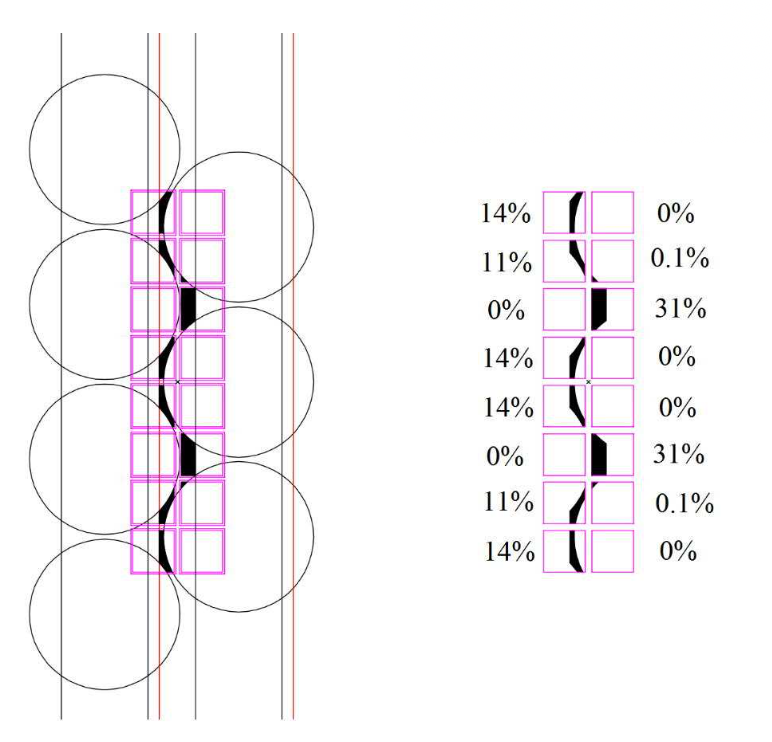}
     \label{LHCbPDMschem}}
   \caption{\subref{PDMpicture}: picture of a photon detector module prototype installed in the \lhcb cavern during the 2018 operations.  \subref{LHCbPDMschem}: schematic view of the setup. The photon detector module, integrating sixteen MaPMTs, is installed behind the HPD plane. Given the HPD staggered configuration, Cherenkov photons reach MaPMTs with the relative intensity reported in the figure.}
  \label{pdmAtLHCb}
\end{figure}

During the last year of data-taking before the Long Shutdown 2, an upgrade photon detector module prototype was installed behind the \richtwo HPD plane, in order to evaluate the behaviour in the \lhc environment (see Fig.~\ref{pdmAtLHCb}). The module was integrated and operated together within the \lhcb detector framework, including services, data acquisition (DAQ) and control systems. During these tests, out-of-time hits, \ie delayed with respect to the expected arrival time of Cherenkov photons relative to a $pp$ collision, were detected, indicating the presence of an unexpected source of noise. The observation of this noise has been made possible by acquiring data in an \lhc collision scheme with isolated bunches, and with a 3\mus-wide acquisition time window, with the aim to synchronise the module with the \lhc clock. 

These noise pulses have been further investigated in the laboratory by means of different experimental setups and light sources, as described in Sec.~\ref{sec:setupDescription}. The characterisation of these signal-correlated events is described in Sec.~\ref{sec:analysis}, and a model to estimate the amount of pile-up effects at high rates is developed and validated in Sec.~\ref{sec:pileUp}. Mitigation strategies have been extensively studied, and the most effective solutions are reported in Sec.~\ref{sec:mitigation}. Finally the impact of this noise on the PID performance of the \rich detectors is assessed using the \lhcb simulation framework as described in Sec.~\ref{sec:PID}.

\section{The \lhcb \rich opto-electronics chain for \lhc Run 3}
\label{sec:mapmtDescription}

The detection of Cherenkov light requires single-photon counting capability in the UV and visible wavelength ranges and at the \lhc bunch crossing rate of 40\mhz. In addition, given the large event multiplicities, position-sensitive photon detectors are necessary to achieve the required Cherenkov angle resolution, distinguish Cherenkov rings close in space and to assign each reconstructed ring pattern to the corresponding charged particle track. 

\begin{figure}[tb]
  \begin{center}
    \includegraphics[width=0.5\linewidth]{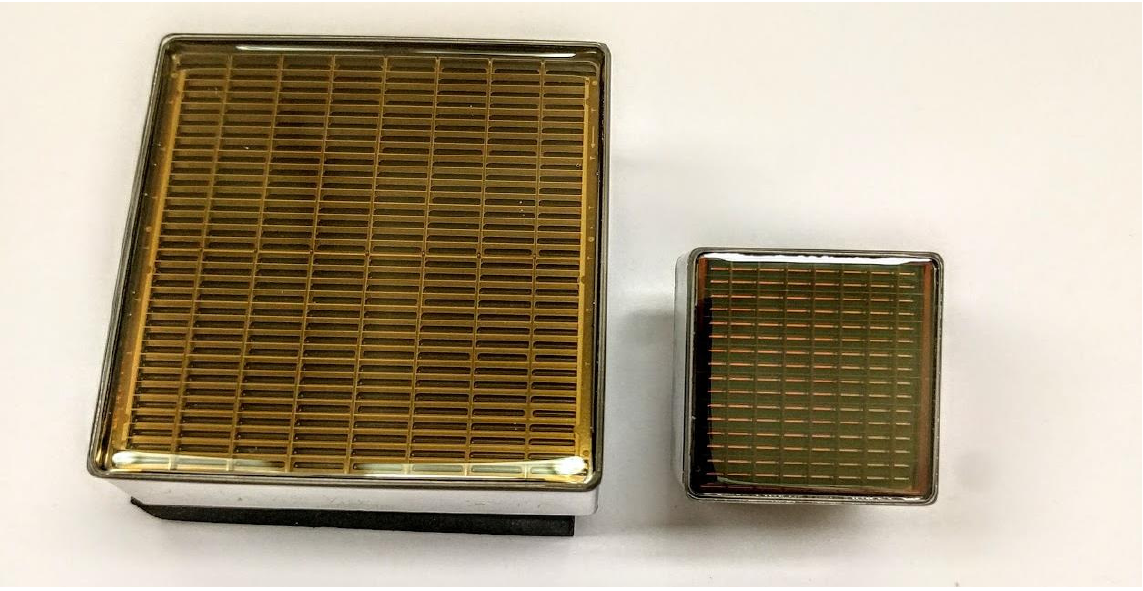}\put(-120,-10){(a)}
    {\includegraphics[width=0.5\linewidth]{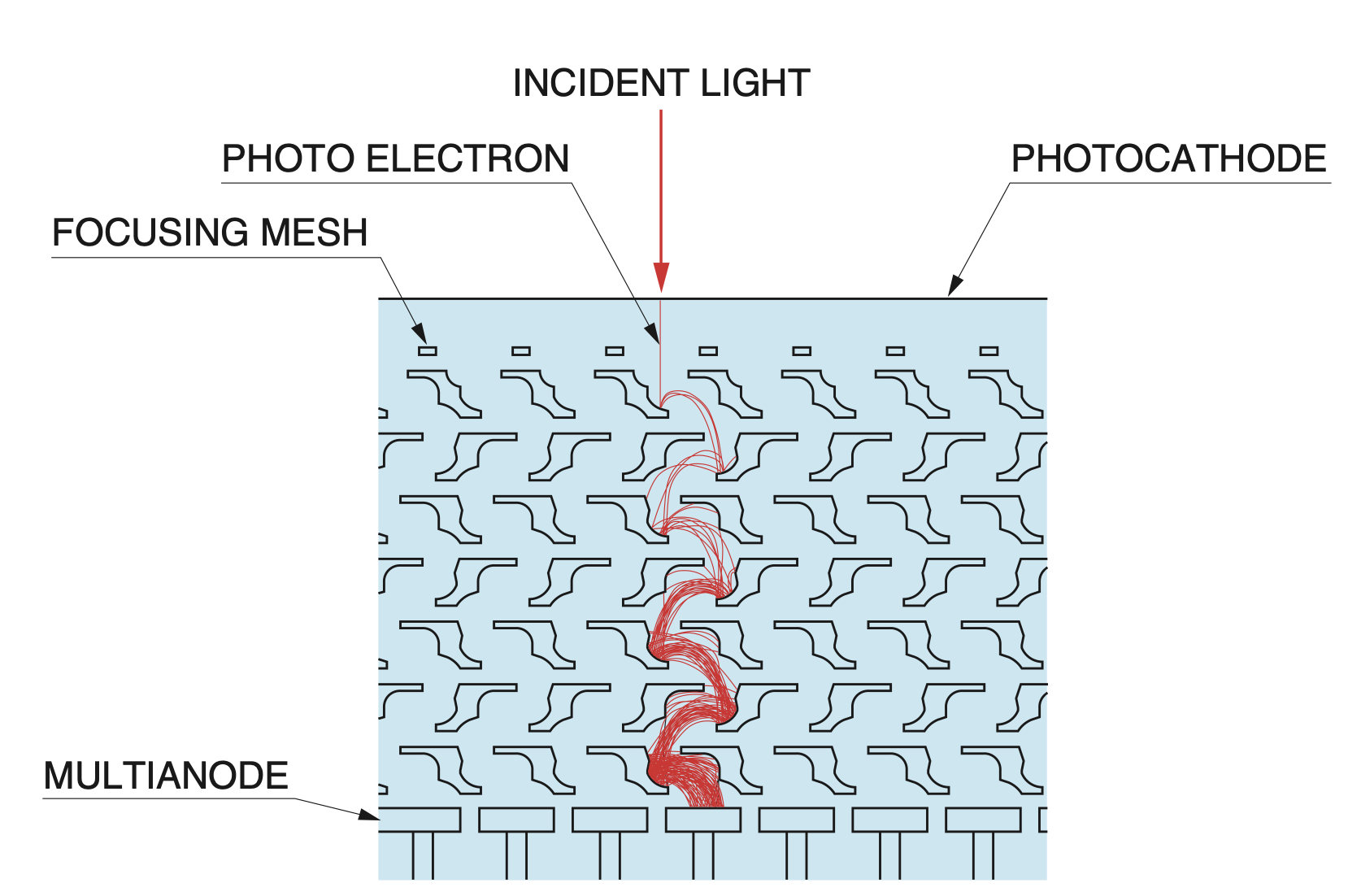}\put(-115,-10){(b)}}
    \vspace*{-0.5cm}
  \end{center}
  \caption{
    (a) The MaPMTs selected for the Run 3 \rich detectors with the 2-inches R13743 on the left and the 1-inch R11265 on the right. (b) Scheme of the electrode structure for the metal channel dynodes and the electron trajectories~\cite{hamamatsuHandbook}.}
  \label{fig:RICH:mapmts}
\end{figure}

A custom version of Hamamatsu R11265 MaPMTs (namely R13742) were selected as photon detectors for \richone and the central region of \richtwo, where the expected occupancy is up to the 30\%.\footnote{A custom version of Hamamatsu R12699 (namely R13743) with a coarser granularity are used in the outer region of \richtwo where the occupancy is about one order of magnitude lower. The noise characterised in this paper is largely absent in such MaPMTs.} These tubes are 26.2$\times$26.2\mma devices composed by $8\times8$ pixels, each of size 2.9$\times$2.9\mma, and are displayed in Fig.~\ref{fig:RICH:mapmts}. About 3000 tubes have been qualified to fulfil several requirements such as an average gain (across all pixels) larger than ${10^6}$, a maximum gain variation of 1:4 among pixels, and a dark count rate lower than 2.5\khz/\cma. The difference between R13742 MaPMTs and R11265 MaPMTs stands on such technical specifications, while other relevant changes are not present in the internal structure of the photon detectors.

The current pulses from the MaPMT anodes are read out by a fast, radiation hard, low power consumption, 8-channel ASIC named CLARO~\cite{Baszczyk:2017fiz}. Each CLARO channel contains an amplifier and a discriminator with programmable threshold.

\begin{figure}[tb]
  \begin{center}
    \includegraphics[width=\linewidth]{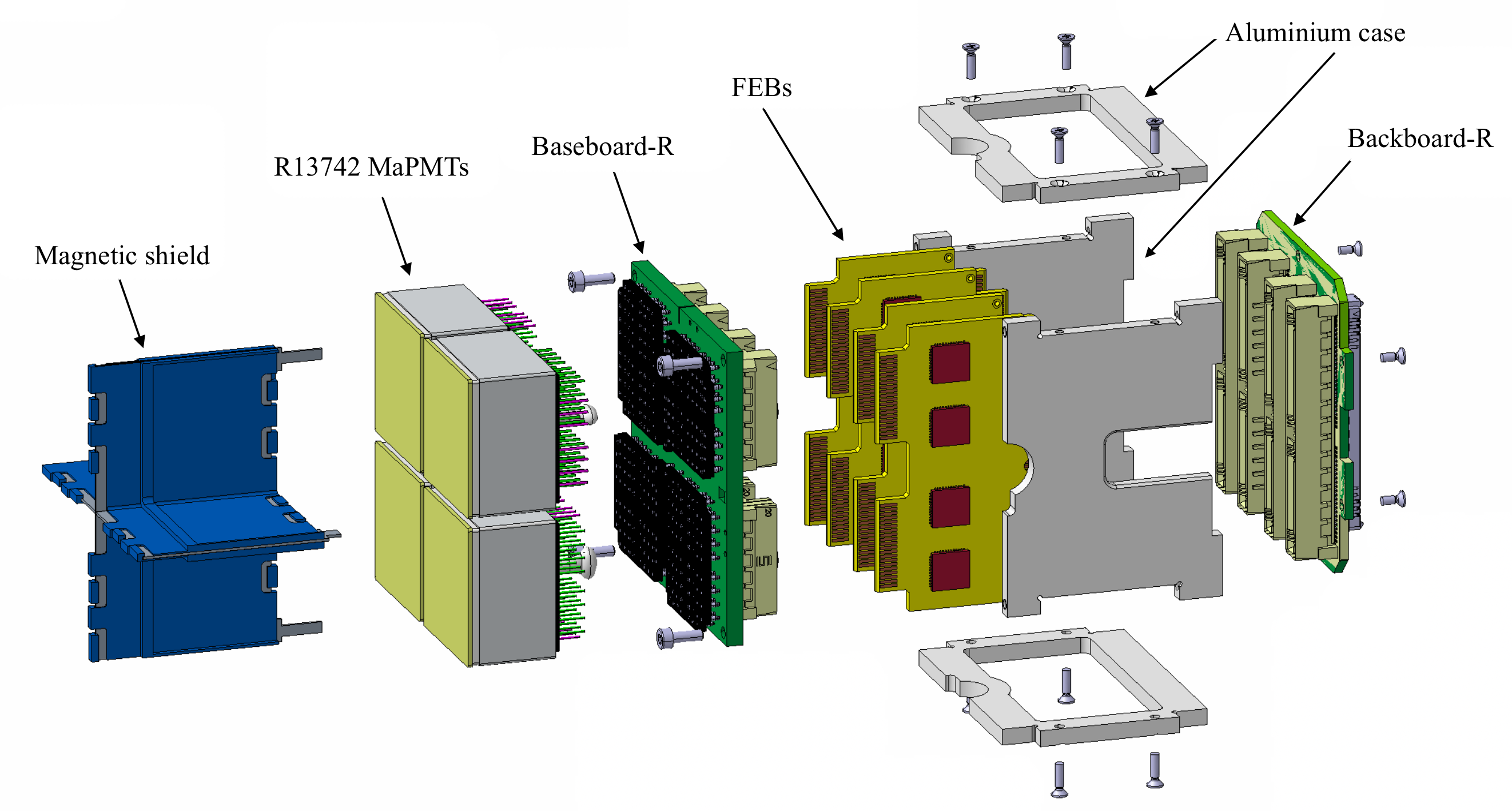}
    \vspace*{-0.5cm}
  \end{center}
  \caption{Exploded view of the elementary cell. The magnetic shield protects the MaPMTs from the effects of the residual magnetic field in the \richone region.}
  \label{fig:RICH:ECRexploded}
\end{figure}

Four MaPMTs and the associated FE electronics are grouped into a unit called Elementary Cell (EC), whose schematic view is reproduced in Fig.~\ref{fig:RICH:ECRexploded}. The MaPMTs are housed by a baseboard, providing the bias to the photon detectors through four custom voltage dividers, and routing the signals from the anodes to the CLARO inputs. Four FE Boards (FEBs) are plugged into the rear side of the baseboard, each FEB hosting eight CLARO ASICs that convert the anode pulses into binary signals. The latter are routed by a backboard to a digital board, namely the Photon Detector Module Digital Board (PDMDB), that captures and formats the binary outputs from the CLARO channels through three Kintex-7 Field-Programmable Gate Arrays (FPGAs). Four ECs and two PDMDBs form a Photon Detector Module (PDM) as shown in Fig.~\ref{PDMpicture}.

\section{Experimental setups}
\label{sec:setupDescription}

Several experimental setups have been used to study and characterise the observed noise, and to ensure that these effects are genuine and do not arise from artefacts of the light source or DAQ system used. 

As mentioned in Sec.~\ref{sec:Introduction} the noise events were first observed while operating a photon detector module prototype in the \lhcb environment. The light source was the Cherenkov photons produced by charged particle tracks in the \cffour gas, \ie the \richtwo radiator, while the DAQ system was a prototype of the pcie40-based back-end system that will be used in the upgraded \lhcb experiment~\cite{Colombo:2018vmp}. The same readout setup is used to perform measurements in the \rich laboratory at \cern. In the latter case, the illumination system is based on a pulsed laser light source synchronised with the DAQ clock and producing single photons at a rate up to 10\mhz. Analogue measurements by probing the anode output of one MaPMT with an oscilloscope are performed as well to exclude that the observed noise was related to digitisation features.

The characterisation of the noise is done on the full MaPMT production by means of a dedicated quality assurance setup allowing to acquire data sets with large statistics. Since most of the studies described in this paper are based on the output of such measurements, the setup is described in detail in the following section.

\subsection{The elementary cell quality assurance setup}
\label{sec:ecqa}

Given the complexity and the large number of ECs (1056 units) needed to equip the \rich focal planes, a Quality Assurance (QA) test setup and protocol have been developed in order to characterise the complete EC production. The ECQA was performed in two sites (Ferrara, Italy and Edinburgh, United Kingdom). Two test stations were deployed in each location. 

\begin{figure}[tb]
   \centering
   \subfigure[]{
     \includegraphics[width=7.5cm, keepaspectratio]{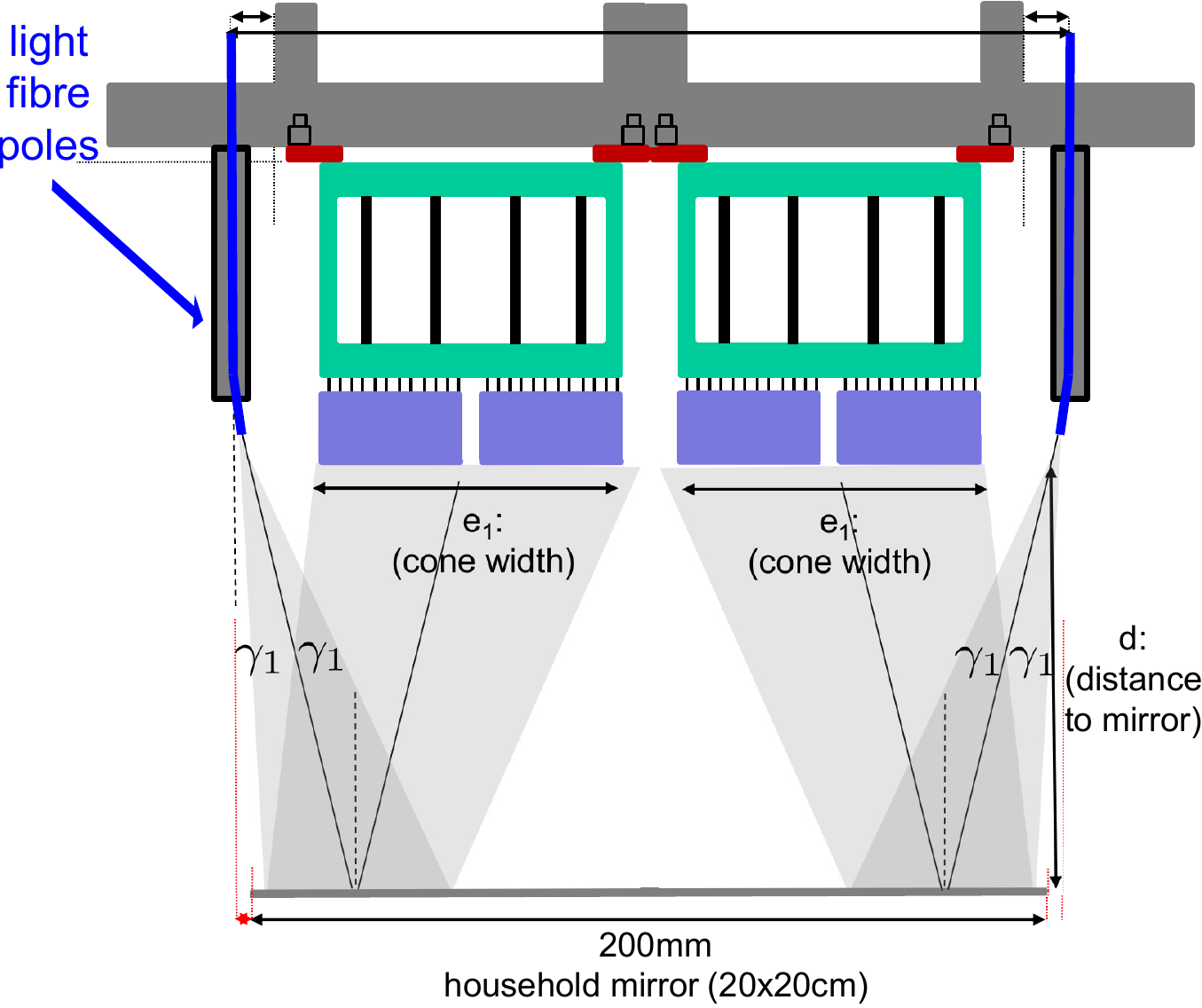}
     \label{ecqaSetupSchematic}}
   \subfigure[]{
     \includegraphics[width=7.5cm, keepaspectratio]{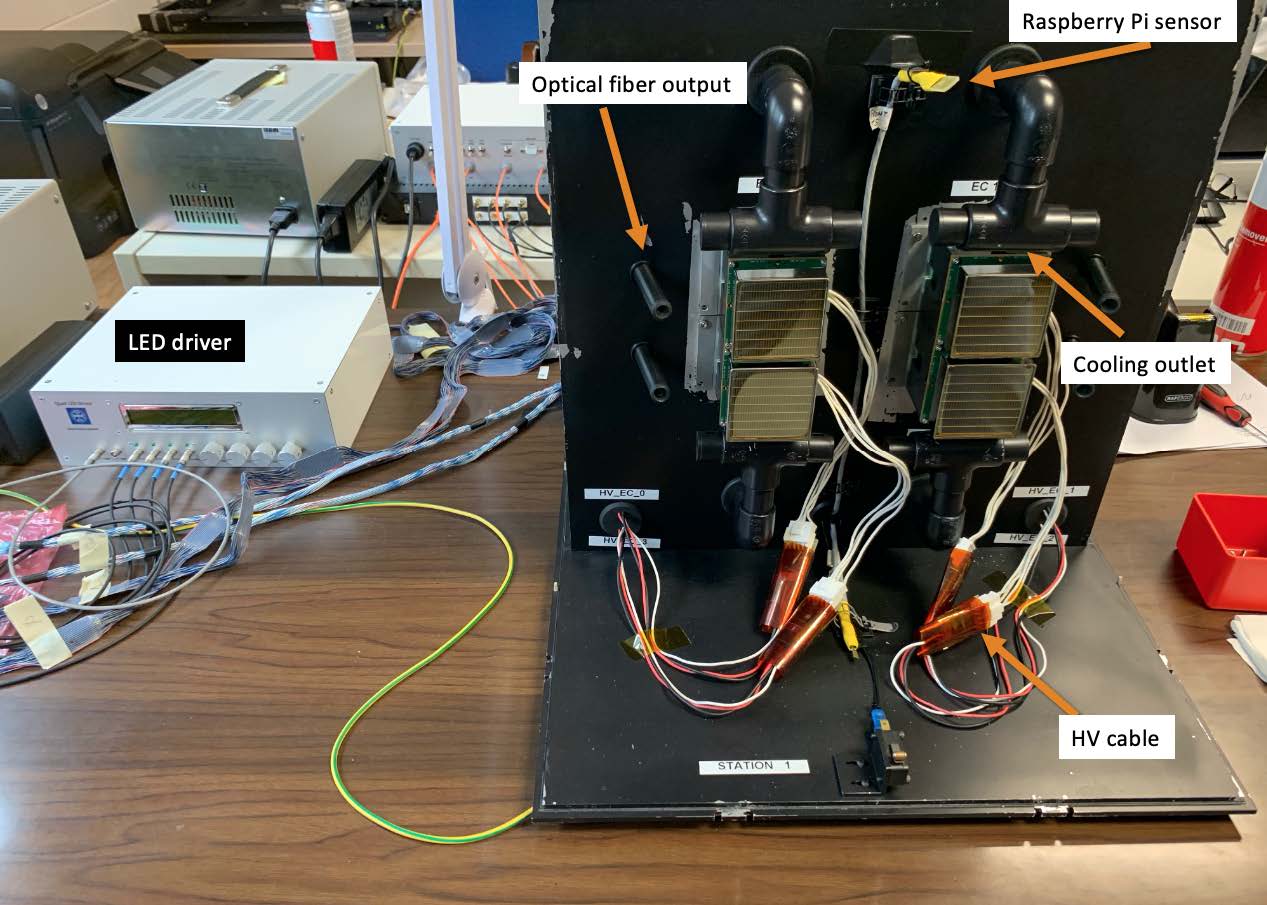}
     \label{ecqaSetupPicture}}
   \caption{Schematic~\subref{ecqaSetupSchematic} and
    picture~\subref{ecqaSetupPicture} of the ECQA setup. The numerical aperture of each fibre is used to achieve a uniform illumination on the photon detectors.}
  \label{Fig:ecqaSetup}
\end{figure}

The test setup consists of a light-tight box that is internally divided in two sections. The back section houses four Digital Boards (DBs), comprising two ALTERA MAX10 FPGAs each. A DB hosts two 12-bit, 8-channel Analogue-to-Digital Converters (ADC) to monitor the CLARO power consumption, together with the current and temperature of the FEBs, baseboards and backboards. The front section contains the ECs and is separated from the back section by a black, light-tight, panel. The illumination system consists of an external blue LED driver. The pulsed LED light is injected in multimode optical fibers that are fed through in the box. Light is reflected off a flat mirror placed inside the box, providing uniform illumination on the MaPMTs. The setup is shown in Fig.~\ref{Fig:ecqaSetup}. 

Environmental temperature and humidity that may affect the MaPMT performance are monitored using a Raspberry Pi and are stabilised by an air-cooling fan system. A system controller, based on the Cyclone V GT FPGA development board, is placed outside the box. It is connected to the DBs through a Universal Asynchronous Receiver/Transmitter (UART) interconnection module and to a PC via a Gbit Ethernet interface. The High Voltage (HV) for the MaPMTs is provided by an ISEG ECH 242 high voltage crate equipped with an ISEG crate controller model CC24 Master6 and an ISEG 8-channels HV power supply board EHS 8020n.

Data acquisition is managed by control software based on Python for the low-level communication with the hardware. A Graphical User Interface (GUI) is implemented in National Instruments LabVIEW. The software controls the HV crate, configures the DBs together with the CLARO ASICs and triggers the LED driver to generate light pulses synchronous to the DAQ clock. The measurements required to validate ECs are carried out automatically by the software and are monitored without operator's intervention. The procedure performs the validation of up to 4 ECs simultaneously for each test station, which corresponds to 16 Hamamatsu R11265 MaPMTs. Raw data are stored to disk and the parameters characterising the behaviour of each single EC are extracted and saved to a database.
 
\section{Characterisation and results}
\label{sec:analysis}

The setup described in Sec.~\ref{sec:ecqa} is
used to record the time spectrum of the noise pulses, hereafter referred to as Signal Induced Noise (SIN) pulses. The SIN acronym has been first introduced in the literature to identify an analogous phenomenon observed in photomultipliers used in LIDAR experiments~\cite{sinLidar}. The recorded time spectra are binned in 25\ns time-slots, and the total DAQ time-window used for data analysis is 6.325\mus. The repetition rate of the laser source, unless stated otherwise, is 100\khz with $10^{6}$ laser pulses per run. The illumination is uniform across the whole photocathode surface. All but those measurements used to study the SIN dependence on the HV, are performed biasing the MaPMTs at 1000 V. 

\subsection{SIN pulses time distribution and probability}
\label{sec:timeDistribution}

\begin{figure}[tb]
   \centering
   \subfigure[]{
     \includegraphics[width=7.5cm, keepaspectratio]{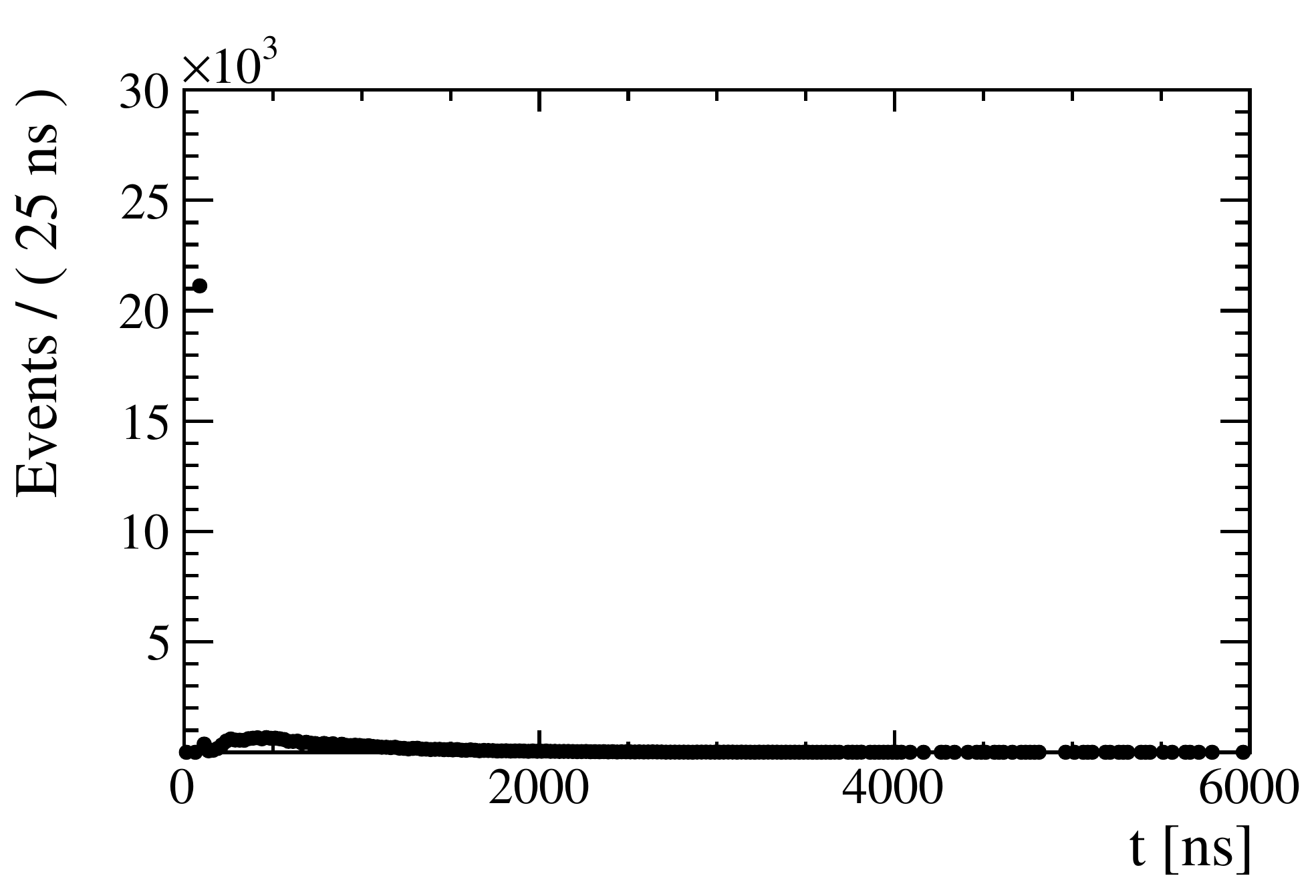}
     \label{sinLinearSpectrum}}
   \subfigure[]{
     \includegraphics[width=7.5cm, keepaspectratio]{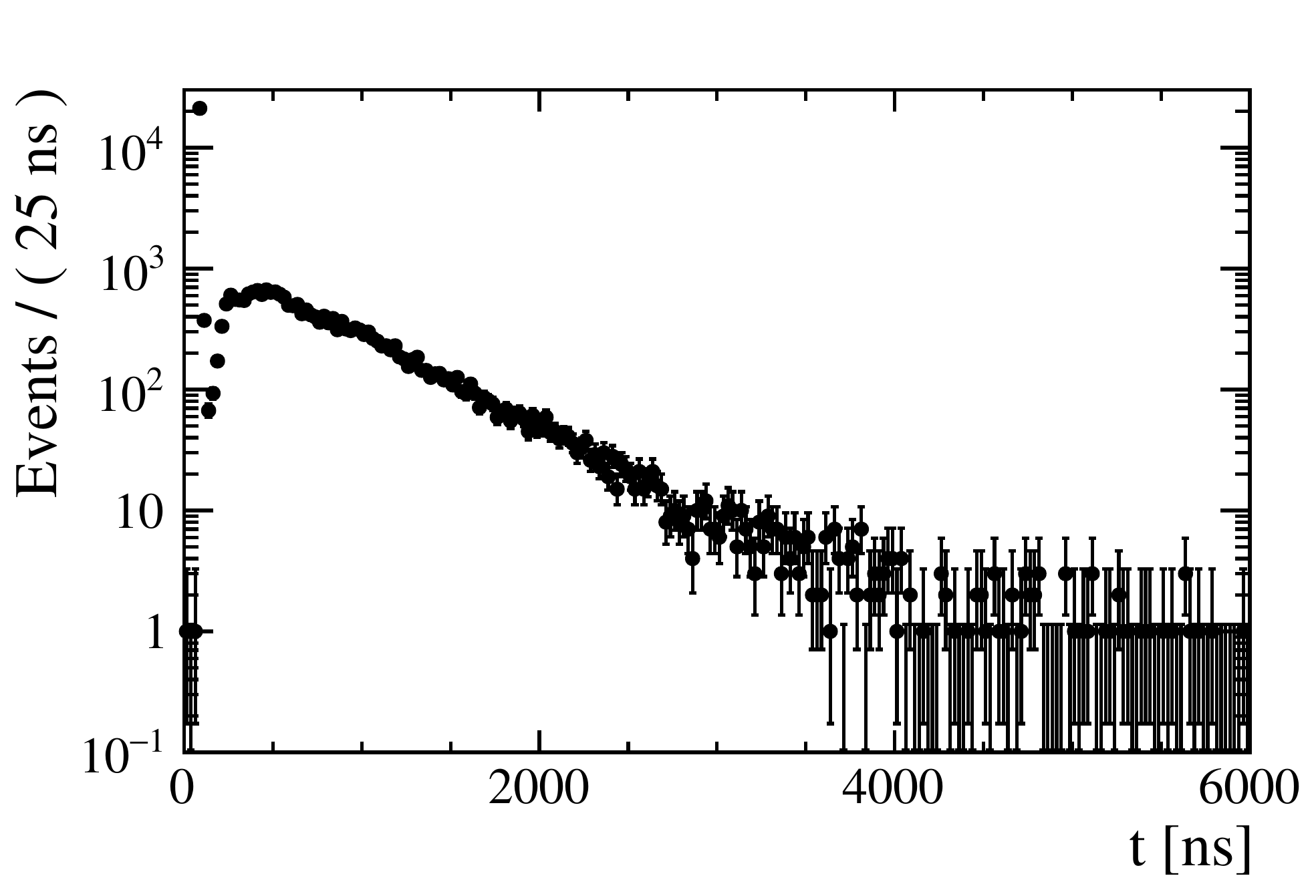}
     \label{sinLogSpectrum}}
   \caption{Time distribution of detected events in \subref{sinLinearSpectrum} linear and
    \subref{sinLogSpectrum} logarithmic vertical-axis scale for a pixel (number 61) affected by
    SIN. The MaPMT under study has serial number FB4500. The corresponding mean number of SIN pulses, treated as a Poisson-distributed variable, is $R_{61}=1.034 \pm 0.010$.}
  \label{SINspectrum}
\end{figure}

A typical spectrum for a pixel affected by SIN is shown in Fig.~\ref{SINspectrum}. The distribution is dominated by the primary laser signal, seen in the leftmost part of the plot as a peak of events populating the 100--125\ns time slots (signal slots). The events after the signal slots are due to SIN pulses.

The probability of generating $n_{\text{sin}}$ noise pulses is treated as Poisson distributed, \ie the processes creating multiple SIN pulses from a single photon converted at the photocathode are considered, in first approximation, as independent~\cite{Haser:2013is}. In vacuum photomultipliers, these processes are typically due to the ionisation
of the residual gas inside the tube caused by the interaction with electrons, generating ion feed-back to the photocathode. But other sources of signal-correlated noise such as light emission are also reported in the literature~\cite{Antochi:2021wik}. The mean number of processes generating SIN pulses is defined as $\mu_{\text{sin}}$. The time profile shown in
Fig.~\ref{SINspectrum} indicates that the number of detected SIN pulses is a function of the time elapsed from the detected primary signal hit.

The probability $P_{m}$ to detect at least one SIN count induced by the primary
signal $m$ time slots after the laser pulse is given by the number of counts $N_m$ in bin
$m$ divided by the number of counts $S$ in the signal slot. Indicating
with $\bar{P}_k$ the probability to detect zero counts in time slot $k$, the following identity can be written: 

\begin{align}
P_{m} = \frac{N_m}{S} = \left ( \prod_{k=1}^{m-1} \bar{P}_{k} \right
  ) (1-\bar{P}_{m}) + \left ( 1-
  \prod_{k=1}^{m-1} \bar{P}_{k} \right ) (1-\bar{P}_{m}) = 1 - \bar{P}_{m}.
\label{sinCountProbabilityGeneral}
\end{align} 

In a first approximation, \ie neglecting the contribution arising from dark counts, the first term in the
sum gives the probability to detect a primary SIN pulse in slot m, while the
second term describes the probability to detect in the same slot higher-order
SIN pulses originating from cascade effects.

Contributions to $P_k$ arise from both dark counts and
SIN pulses. Dark counts, obeying Poisson statistics and
occurring with a mean number of counts $\mu_d$ in a $25 \ns$ bin,
contribute to $P_{m}$ independently of the presence of
signal. $\bar{P}_{\text{d}} = e^{-\mu_\text{d}}$ and
$\bar{P}_{\text{sin},m} = e^{-\mu_\text{sin}(m)}$ are the probabilities to have zero dark
counts and SIN pulses, respectively. The probability to detect zero counts in time slot $m$ is therefore given by 

\begin{align}
\bar{P}_{m} = \bar{P}_{\text{d}} \cdot \bar{P}_{\text{sin},m}.
\label{noCountsProb}
\end{align} 

The time profile of the
SIN pulse probability is introduced through a dependence of the
mean number of processes on the slot under
consideration. Considering explicitly and separately dark counts and SIN pulse
contributions, and using Eqs.~\ref{sinCountProbabilityGeneral} and~\ref{noCountsProb}, the probability to detect at least one SIN pulse in
slot $m$ after the signal can be written as

\begin{align}
P_{\text{sin},m} = \left ( \frac{N_m}{S} -1 \right )e^{\mu_d} + 1 .
\label{sinCountProbability}
\end{align} 

Knowing the dark count rate from the MaPMT QA characterisation, it is possible to determine the SIN pulse probability in each slot $m$. 

The probability $P_{\text{sin}}$ to detect at least one SIN pulse in the full time range where SIN pulses occur is given by 

\begin{align}
P_{\text{sin}}  = 1 - \prod_{m=1}^{M} (1- P_{\text{sin},m})
\label{totApProb}
\end{align} 

where the upper limit of the time range
is defined as $25 \ns \times M$. The typical value of maximum dark count rate per pixel is $\nu_d = 1 \khz$, resulting
in a maximum mean number of dark counts per $25 \ns$ bin of $\mu_d
\sim 10^{-5}$. The dark counts contribution can therefore be safely
neglected, since it is at least one order of magnitude lower than the
intensity of SIN pulses. With this approximation, and given that $N_m \ll S$, Eq.~\ref{totApProb}
can be written as 

\begin{align}
P_{\text{sin}}  = 1 - \prod_{m=1}^{M} (1- P_{\text{sin},m}) \approx 1 - \prod_{m=1}^{M}  \left ( 1- \frac{N_m}{S}
  \right ) \approx 1 - \prod_{m=1}^{M}  e^{-\frac{N_m}{S}}  = 1 - e^{-\frac{B}{S}},
\label{totApProbApprox}
\end{align} 

where $B = \sum_{m=1}^{M}
N_m$ is the total number of hits after the signal. The 
 mean number of SIN pulses $\mu_{\text{sin}}$ is therefore
given to a good approximation by the ratio $B/S$, and can be inferred
from the time distribution like the one shown in
Fig.~\ref{SINspectrum}. For this reason the figure of merit used to
assess the effect of SIN on a pixel-by-pixel basis is the ratio
$\mu_{\text{SIN}}^{p}=B_p/S_p$, representing the mean number of SIN pulses for the
pixel $p$ under study. 

\begin{figure}[tb]
   \centering
   \subfigure[]{
     \includegraphics[width=7.5cm, keepaspectratio]{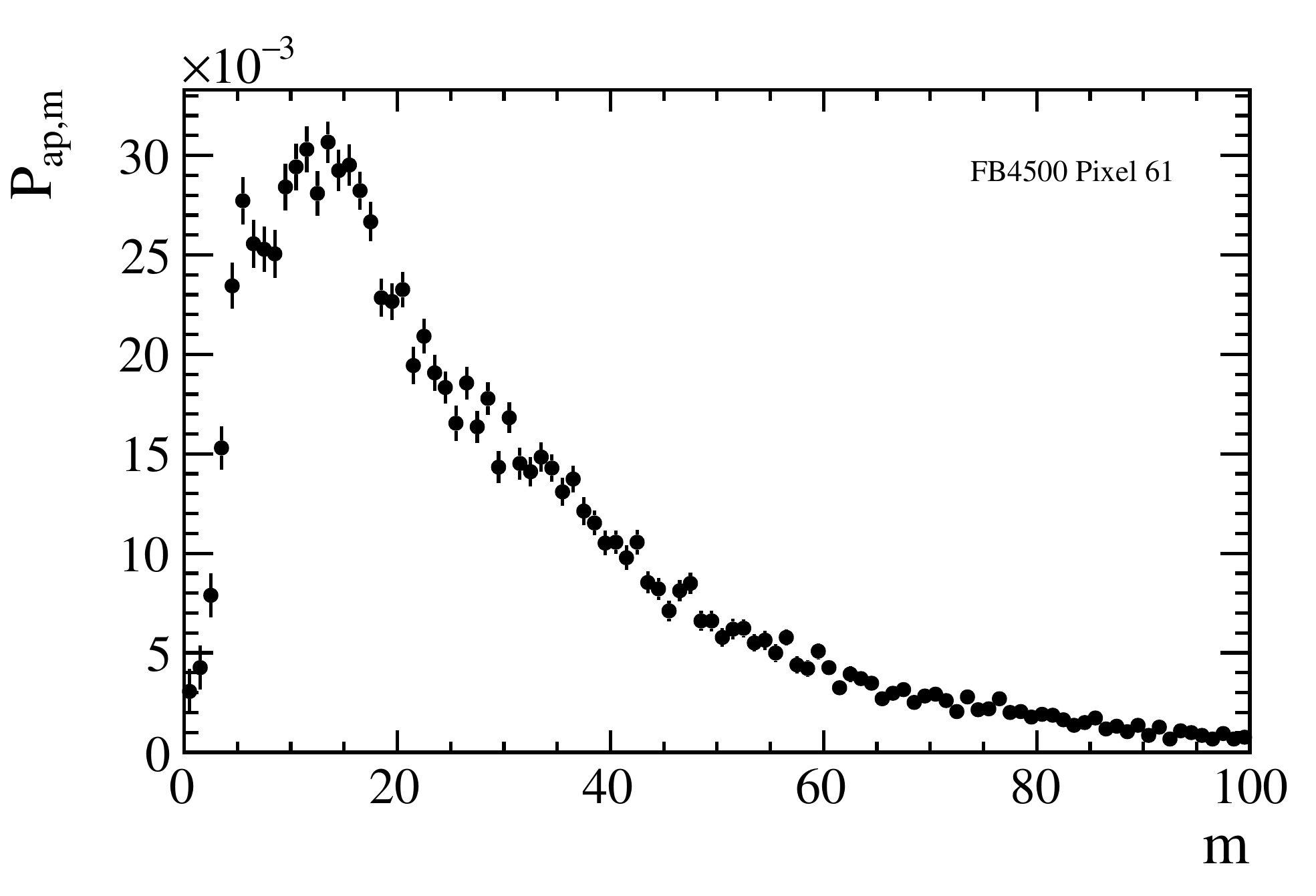}
     \label{pApSpectraFB4500}}
   \subfigure[]{
     \includegraphics[width=7.5cm, keepaspectratio]{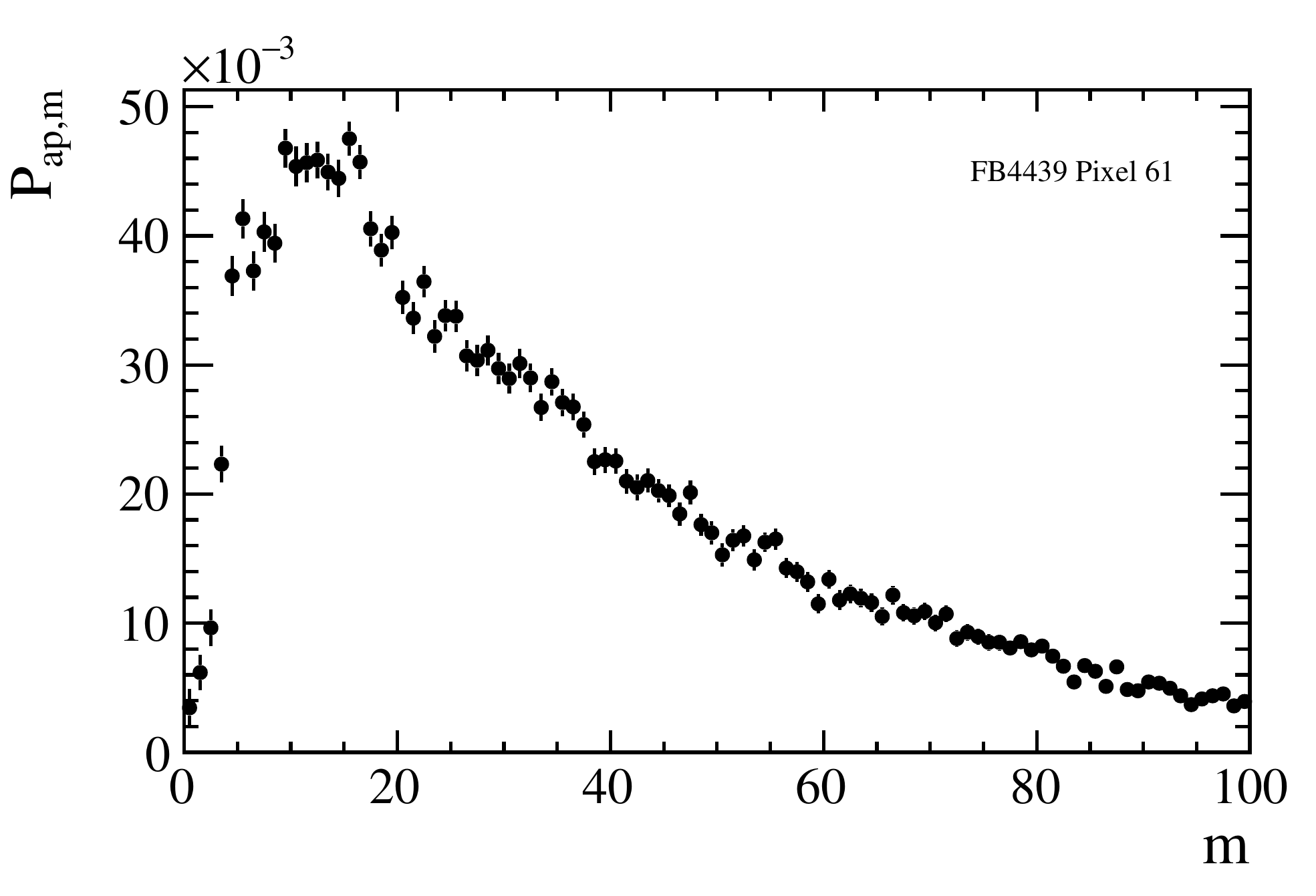}
     \label{pApSpectraFB4439}}
   \caption{SIN pulse probability up to $m=100$ time slots after the signal, corresponding to $2.5 \mus$. The same pixel (number 61) is
     shown for two different MaPMTs, to highlight the qualitative similarity between different devices. The distribution in Fig.~\ref{pApSpectraFB4500} is extracted from the spectrum shown in Fig.~\ref{SINspectrum}. Slot number 1 in the SIN probability distributions corresponds to slot number 6 in the whole distribution of Fig.~\ref{SINspectrum}, \ie the signal slot is not shown here.}
  \label{pApSpectra}
\end{figure}

The mean number of SIN pulses is different for each pixel within the MaPMT under study, and for the same pixel between different
MaPMTs, as shown in Fig.~\ref{pApSpectra}, where the distributions only show background events. However, a common pattern
of pixels affected by SIN is found, and is described in
Sec.~\ref{sec:localisation}. The same figure of merit is used to search for correlations with parameters such as gain and dark count rate, and to study the dependence on the supplied high voltage, the latter being described in Sec.~\ref{sec:hvDependence}. 

\subsection{Localisation of SIN and correlation with MaPMT parameters}
\label{sec:localisation}

\begin{figure}[tb]
   \centering
   \subfigure[]{
     \includegraphics[width=7.5cm, keepaspectratio]{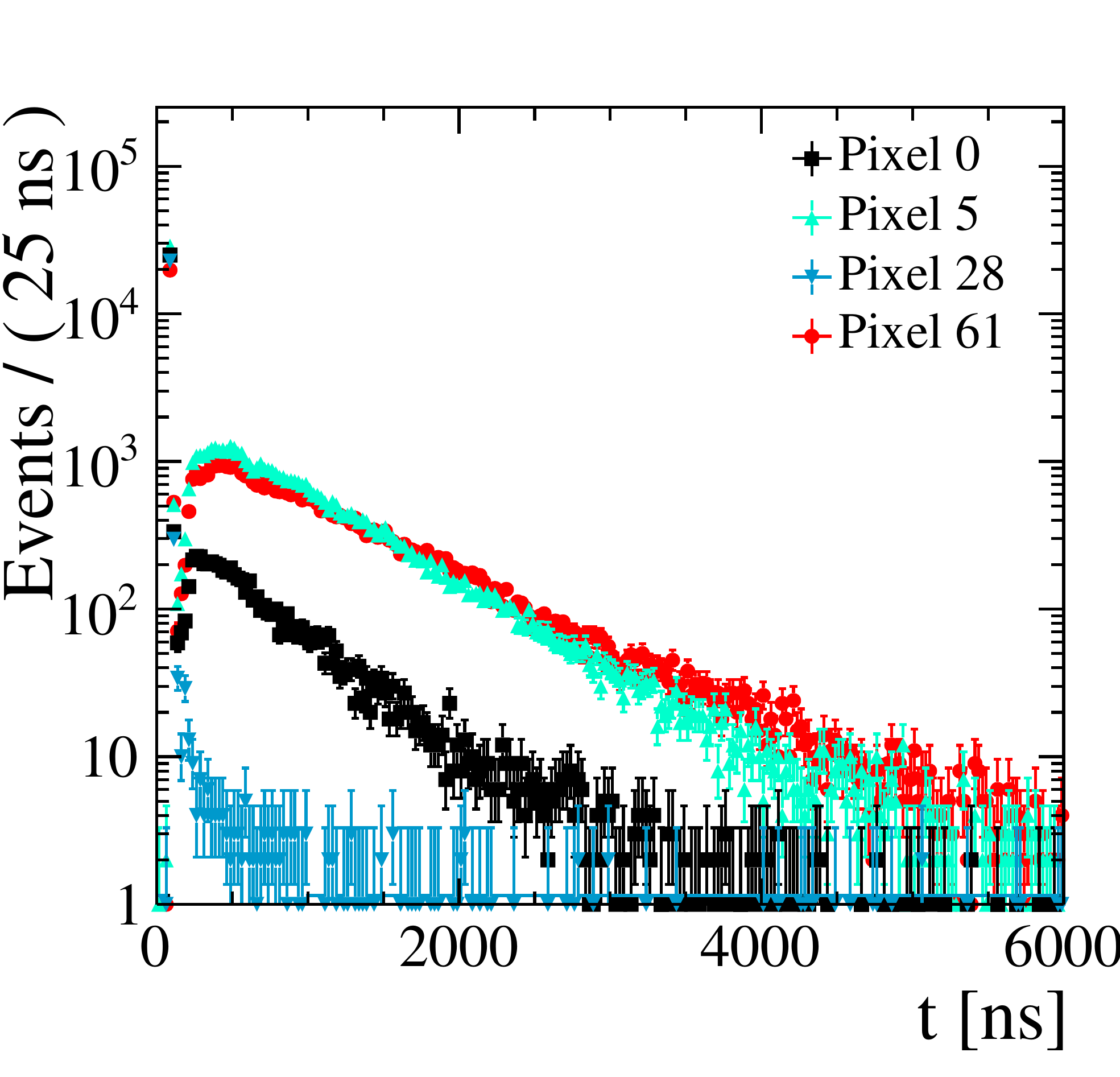}
     \label{sinSpectraComparison}}
   \subfigure[]{
     \includegraphics[width=7.5cm, keepaspectratio]{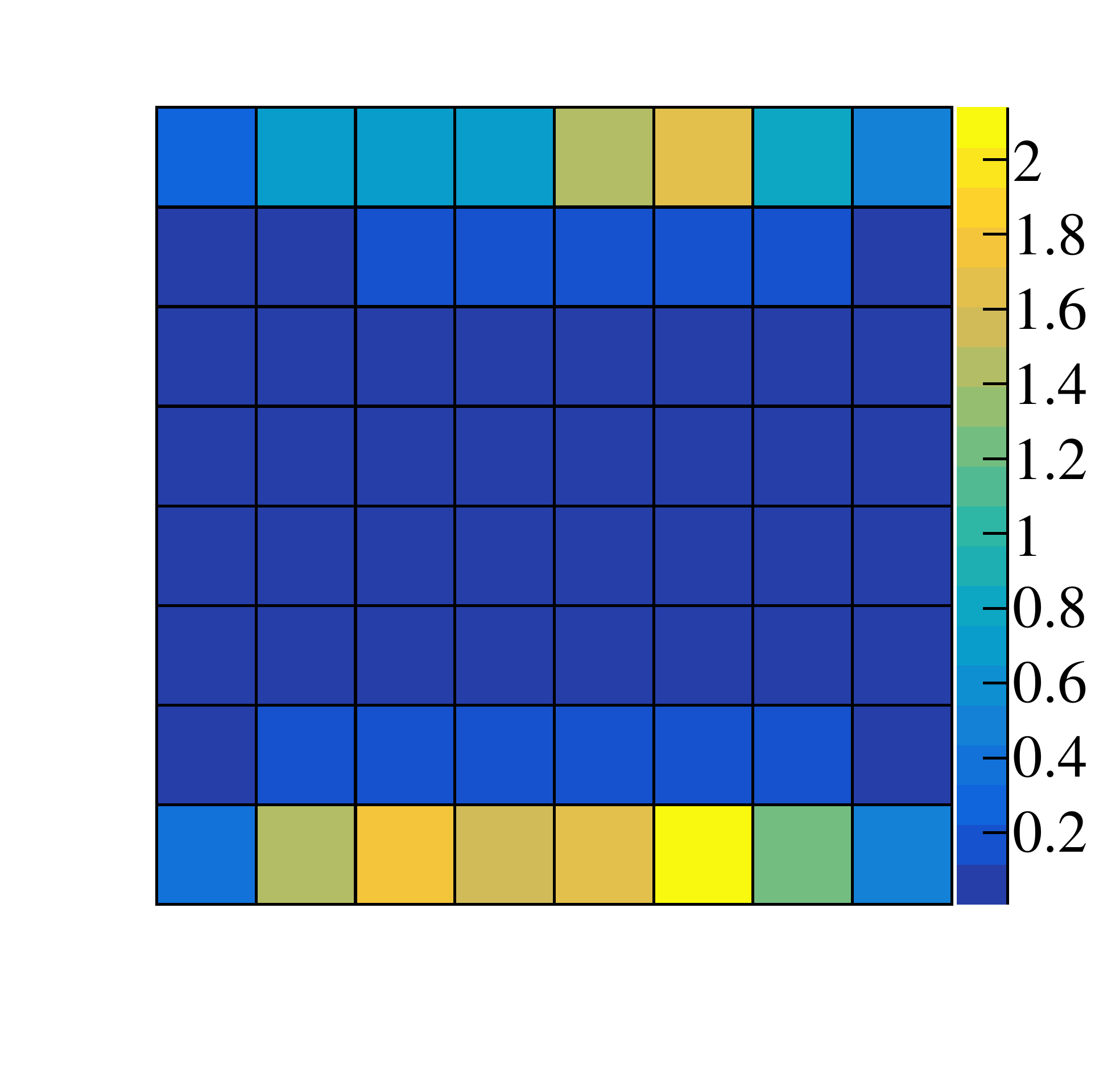}
     \label{localisationMap}}
   \caption{\subref{sinSpectraComparison}: time-spectra in
     logarithmic vertical scale for four pixels of tube FB4439. \subref{localisationMap}: map of the
     mean number of SIN pulses ($\mu_{\text{SIN}}$) for all the pixels of tube FB4439 (pixel-0 is
    top-left, pixel-7 is top right and pixel-63 is bottom-right).}
  \label{SINmap}
\end{figure}

Time spectra such as those shown in Fig.~\ref{SINspectrum} were acquired for all pixels of several
MaPMTs. The time spectra for four pixels and the map of the $\mu_{\text{SIN}}^{p}$
values for the MaPMT with serial number FB4439,
representative of the average quality of R11265 MaPMTs, are reported in
Fig.~\ref{SINmap}.  

The time spectrum for pixel 28 shows that this pixel is essentially unaffected by SIN,
since the events after the signal, beside a negligible component of fast after-pulses, are consistent with dark counts. The
time spectra for the other three pixels clearly show the
effect of SIN, with different intensities that depend on the pixel position within the MaPMT. 

\begin{table}[h]
  \caption{Mean number of SIN pulses estimated for four MaPMTs. For comparison, the first line in the table corresponds to FB4439, showing the numerical ratios computed from the time distributions and map displayed in Fig.~\ref{SINmap}.}
\begin{center}
\begin{tabular}{l D{,}{\pm}{-1} D{,}{\pm}{-1} D{,}{\pm}{-1} D{,}{\pm}{-1}}
    \midrule
    MaPMT & \multicolumn{1}{c}{$\mu_{\text{sin}}^{0}$} &
                                               \multicolumn{1}{c}{$\mu_{\text{sin}}^{5}$}
  & \multicolumn{1}{c}{$\mu_{\text{sin}}^{28}$} & \multicolumn{1}{c}{$\mu_{\text{sin}}^{61}$}\\ 
    \midrule
    FB4439 & 0.2457\hspace{1mm},\hspace{1mm}0.0035 &
                                            1.695\hspace{1mm},\hspace{1mm}0.013 & 0.0109\hspace{1mm},\hspace{1mm} 
                                                       0.0007 & 
                                                                 2.139\hspace{1mm},\hspace{1mm}0.018\\
   FB2294 & 0.0745\hspace{1mm},\hspace{1mm}0.0012 &
                                                     0.3715\hspace{1mm},\hspace{1mm}0.0029 & 0.0076\hspace{1mm},\hspace{1mm} 
                                                       0.0004 & 
                                                                 0.491\hspace{1mm},\hspace{1mm}0.003\\
   FB2312 & 0.132\hspace{1mm},\hspace{1mm}0.0017 &
                                                     0.747\hspace{1mm},\hspace{1mm}0.004 & 0.0081\hspace{1mm},\hspace{1mm} 
                                                       0.0004 & 
                                                                 2.231\hspace{1mm},\hspace{1mm}0.011\\
   FB4500 & 0.266\hspace{1mm},\hspace{1mm}0.004 &
                                                     1.398\hspace{1mm},\hspace{1mm}0.012 & 0.0103\hspace{1mm},\hspace{1mm} 
                                                       0.0007 & 
                                                                 1.034\hspace{1mm},\hspace{1mm}0.010\\
    \midrule
  \end{tabular}
\end{center}
\label{tableLocalisation}
\end{table}

The localisation of SIN can be visualised by computing the $\mu_{\text{SIN}}^{p}$ map, reproduced in
Fig.~\ref{localisationMap}. The SIN effect is essentially concentrated in the top and bottom rows
of the MaPMT. After computing the same map for different MaPMTs, it is observed that the
localisation described above is a general feature of SIN, but the
absolute values of $\mu_{\text{SIN}}^{p}$ differ between different units, as reported
in Table~\ref{tableLocalisation}.

\begin{figure}[tb]
   \centering
      \subfigure[]{
         \includegraphics[width=7.5cm, keepaspectratio]{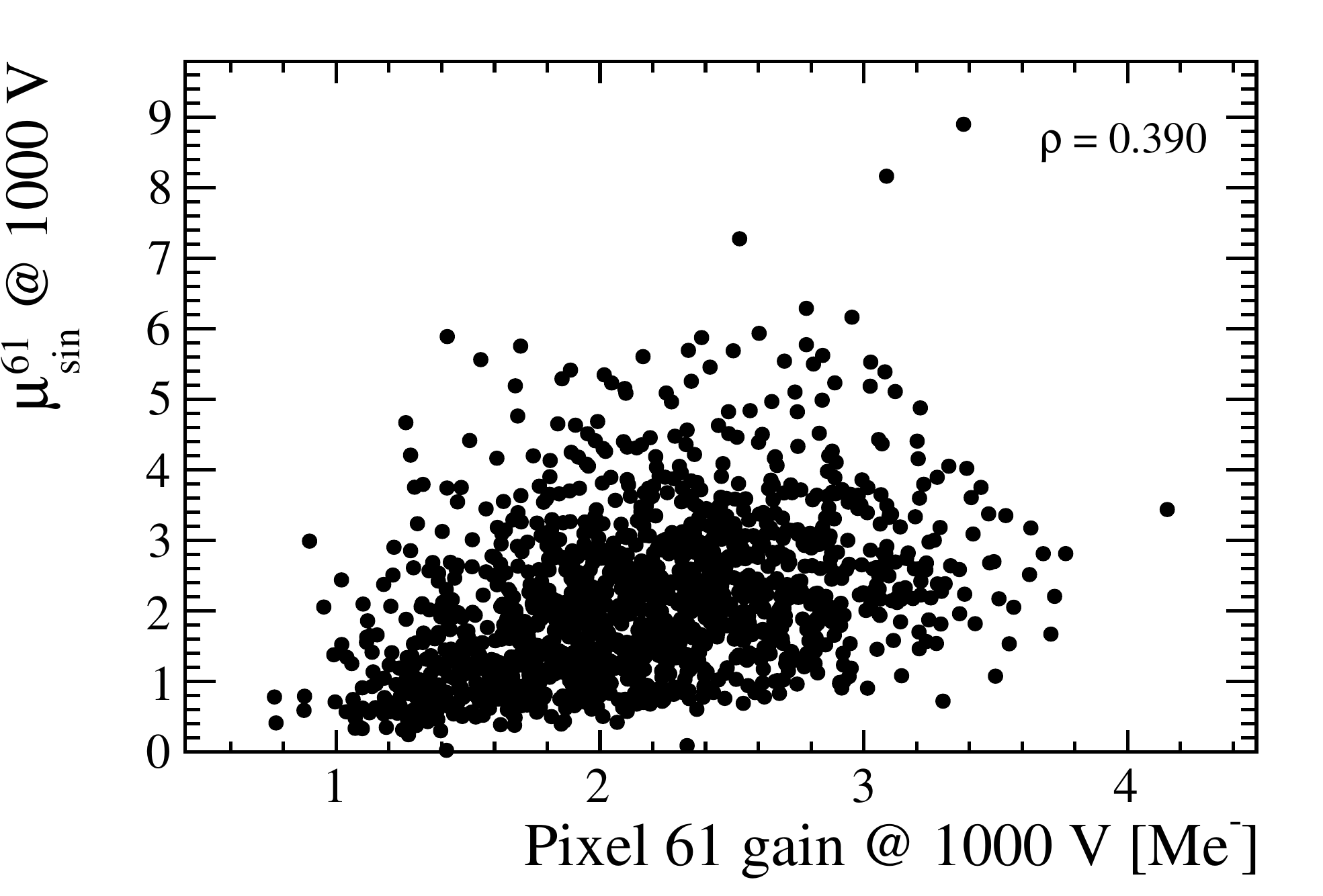}
     \label{correlationGain1000}}
     \subfigure[]{
         \includegraphics[width=7.5cm, keepaspectratio]{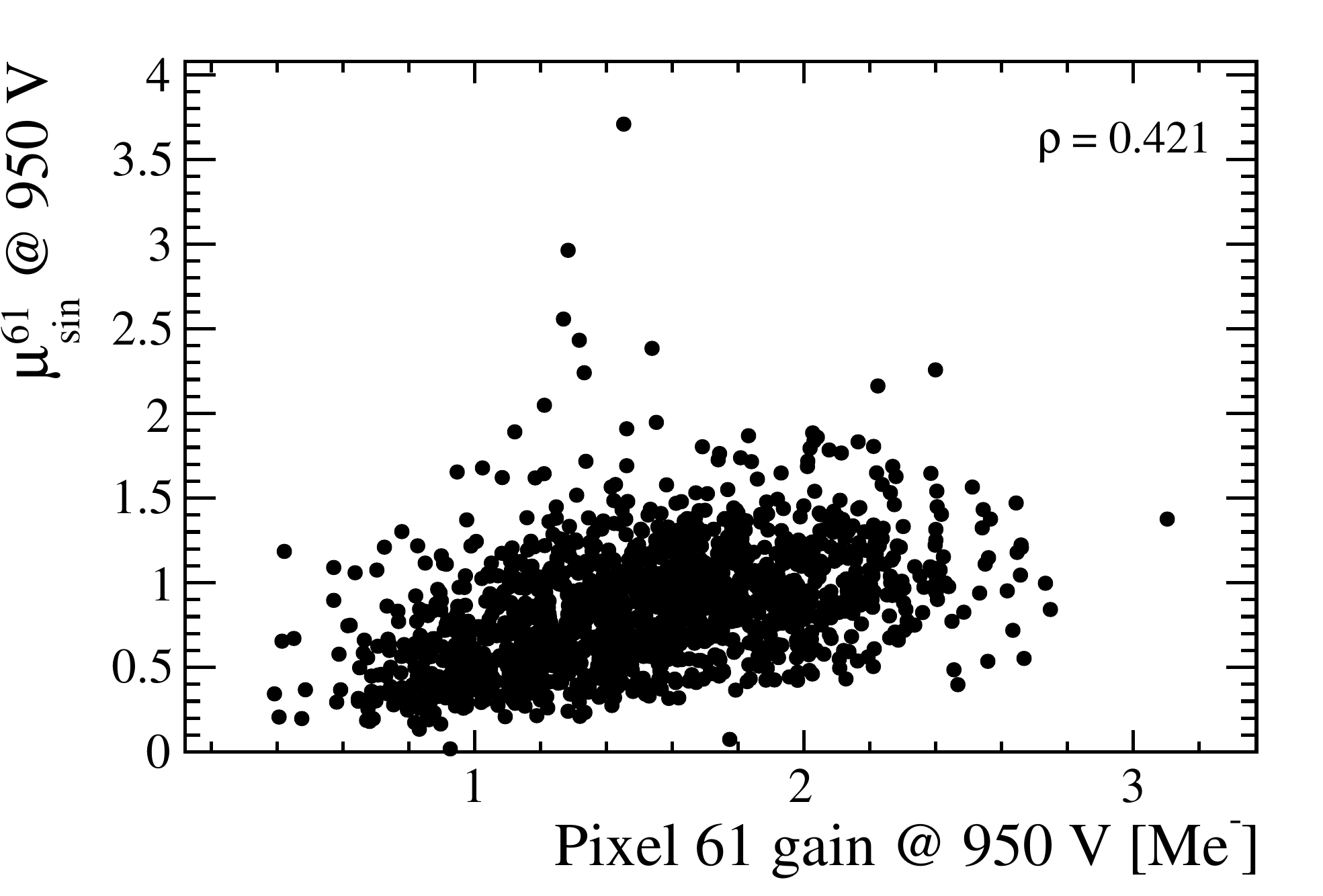}
     \label{correlationGain950}}
     \subfigure[]{
         \includegraphics[width=7.5cm, keepaspectratio]{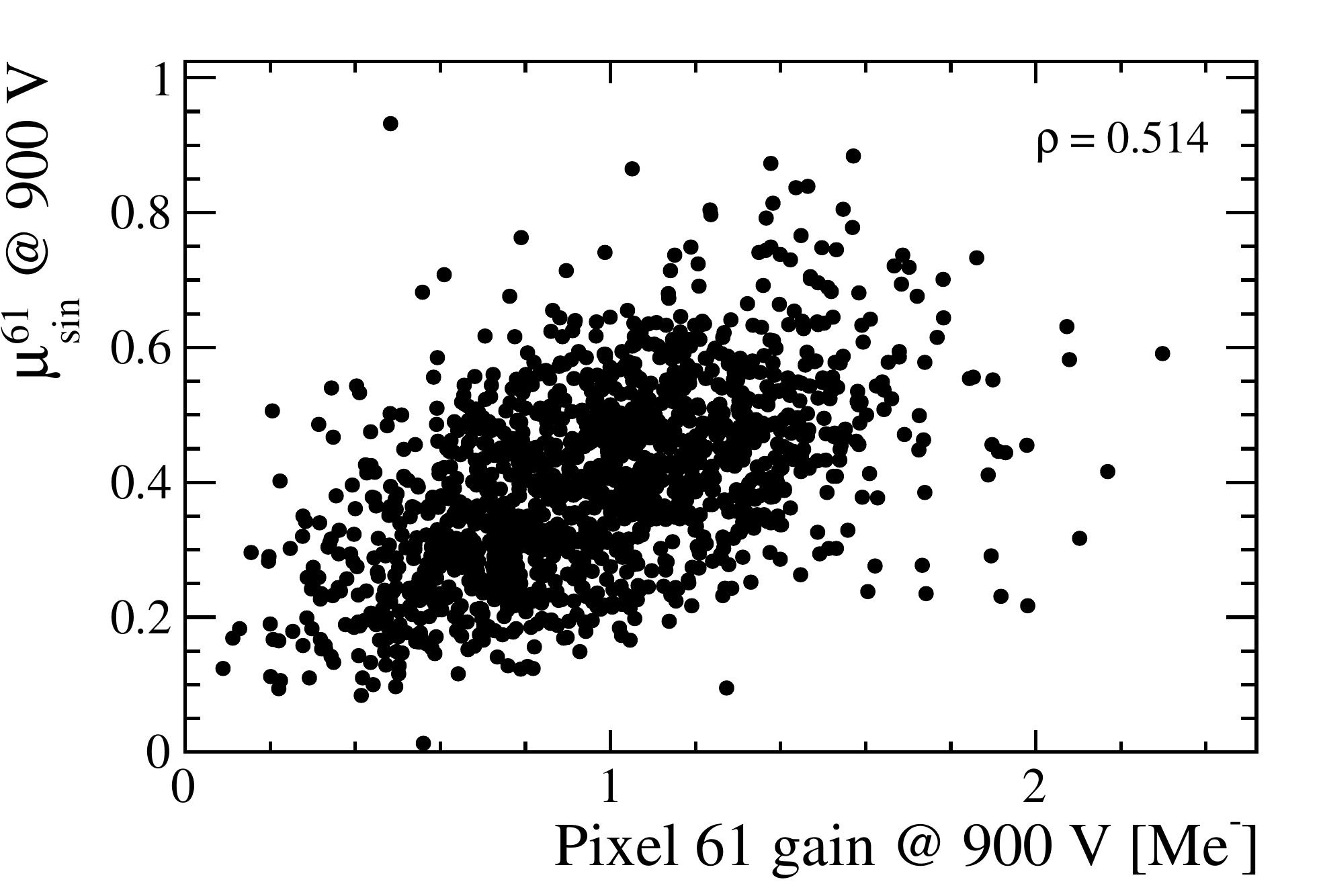}
     \label{correlationGain900}}
     \subfigure[]{
         \includegraphics[width=7.5cm, keepaspectratio]{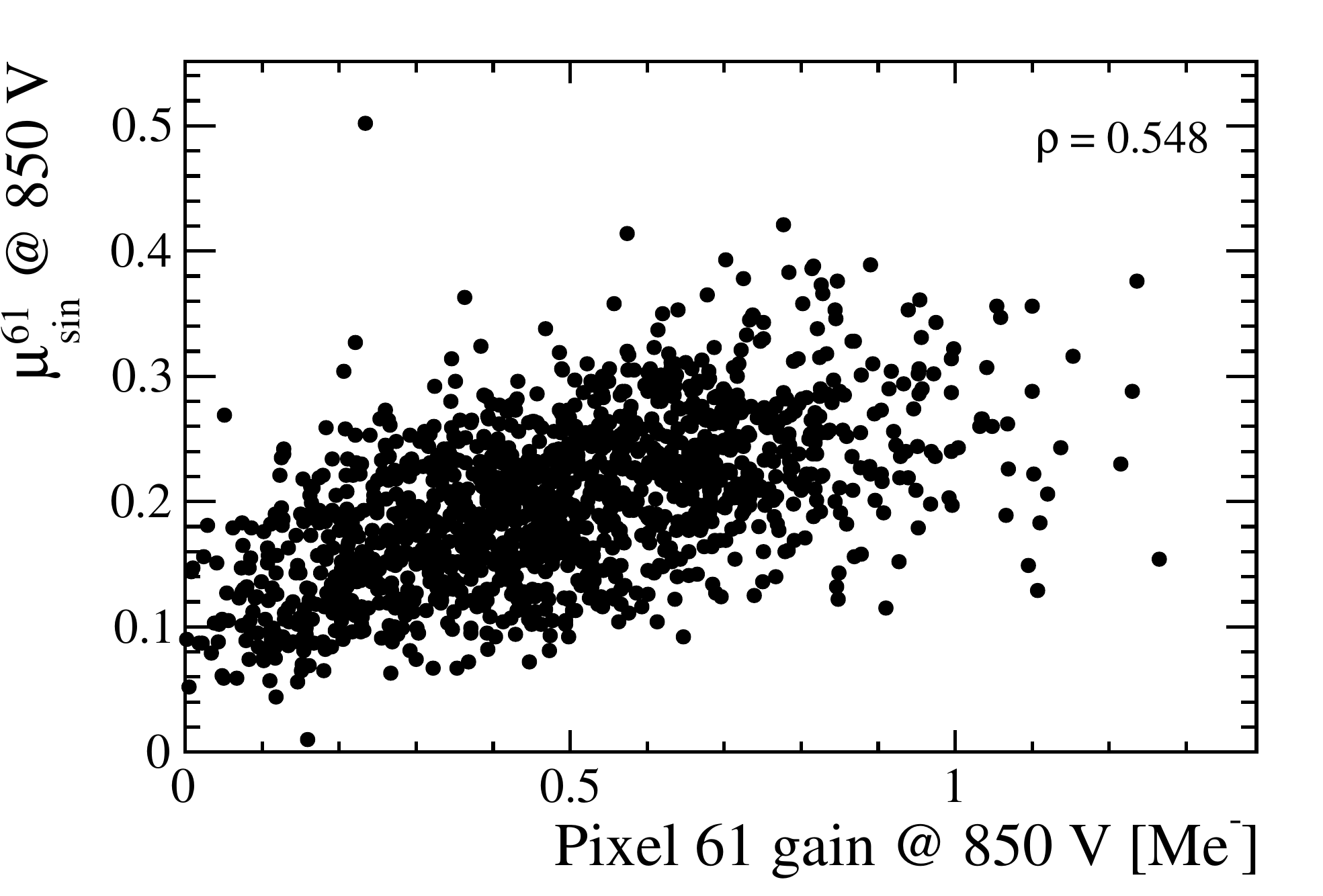}
     \label{correlationGain850}}
   \caption{Correlation between the pixel gain at  \subref{correlationGain1000}~1000 V,
     \subref{correlationGain950}~950 V,
     \subref{correlationGain900}~900 V and \subref{correlationGain850}~850 V and the mean number of SIN events at the corresponding HV.}
  \label{correlationPlotsSecond}
\end{figure}

The correlation between the mean number of SIN pulses and the average gain, blue sensitivity index, average dark current and pixel dark counts for pixel 61 of 1400 MaPMTs of series FB was studied. No clear correlation is observed with respect to these variables. A weak correlation against the pixel gain, increasing at lower high-voltages, is instead visible in Fig.~\ref{correlationPlotsSecond}. This correlation coefficient is determined for each pixel of these 1400 MaPMTs, in order to produce the corresponding correlation maps. The increasing $\mu_{\text{SIN}}$-to-gain correlation coefficient with lower high-voltages matches the MaPMT regions affected by SIN.

Two MaPMTs with a large SIN rate have been subjected to a high illumination rate for a period approximately equivalent to the total integrated charge expected at the end of Run 3. The results of such measurements show that there is no degradation in the MaPMT properties such as the quantum efficiency, while the gain decrease shows no connection between ageing effects and SIN. 

Other tests have been performed by using different voltage dividers, allowing to change the potential difference either between the photocathode and the first dynode or between the last dynode and the anode. No significant changes in the mean number of SIN events and in the time distributions were observed.

\subsection{High-voltage dependence and peaking structures}
\label{sec:hvDependence}

\begin{figure}[tb]
   \centering
   \subfigure[]{
     \includegraphics[width=7.5cm]{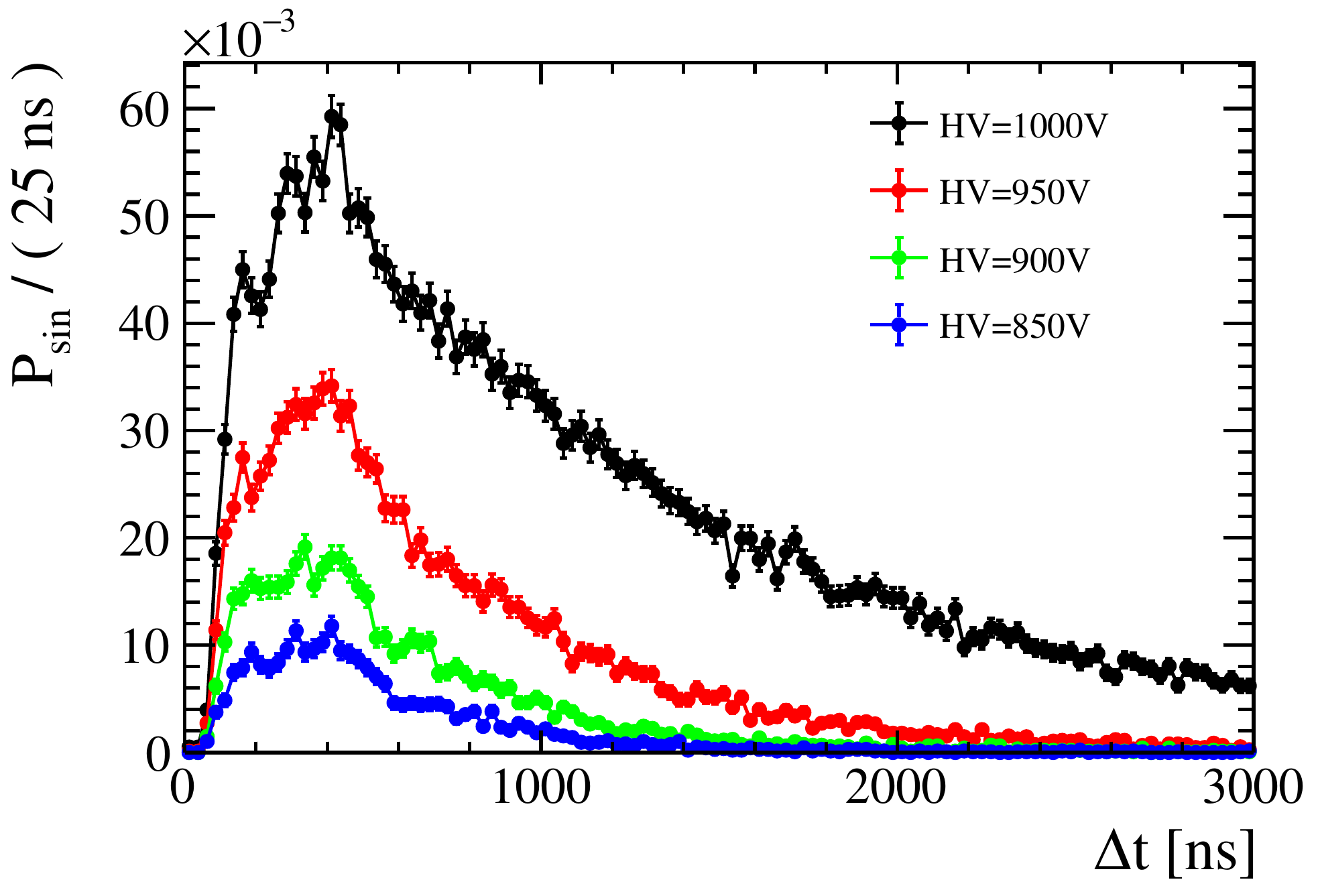}
     \label{HVdependencePlot}}
   \subfigure[]{
     \includegraphics[width=7.5cm]{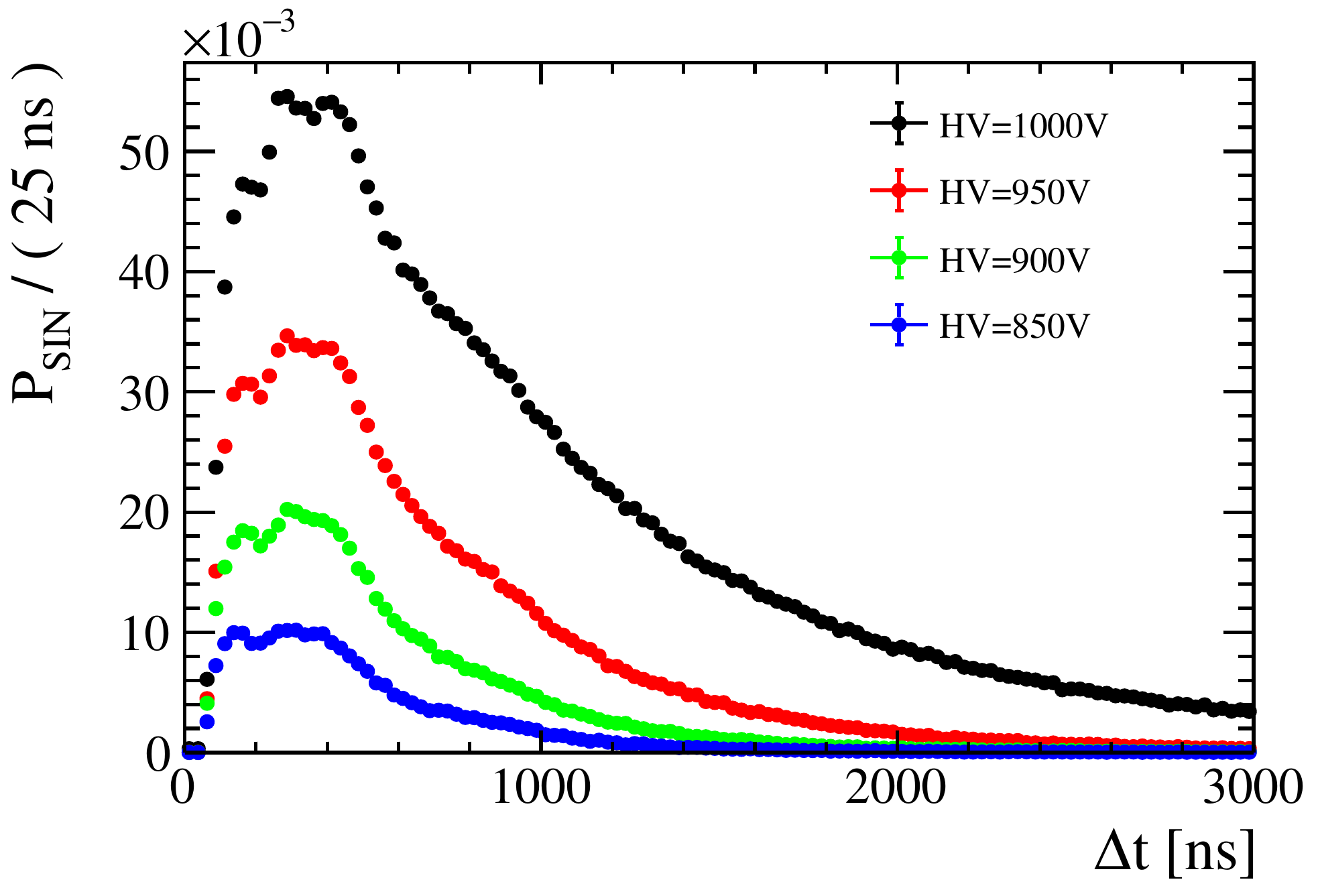}
     \label{multipleHVdependence}}
   \caption{\subref{HVdependencePlot}: comparison between the SIN time distributions at different HV values for anode 61 of one MaPMT. \subref{multipleHVdependence}: Comparison between the SIN time distributions at different HV values for anode 61 averaged over twenty MaPMTs. The characteristics time structure can be seen.}
  \label{HVdependencePlots}
\end{figure}

The SIN probability as a function of the time elapsed from the signal slot and for different applied high voltages is shown in
Fig.~\ref{HVdependencePlot} for pixel 61 of one MaPMT. As explained in Sec.~\ref{sec:timeDistribution}, each bin content corresponds to $N_m/S$, \ie histogram entries are normalised to the signal counts. To highlight the
common time characteristics, the per-bin SIN probability
distribution averaged over a sample of randomly selected 20 MaPMTs is reported in
Fig.~\ref{multipleHVdependence}, for the same pixel 61. 

\begin{figure}[tb]
   \centering
   \subfigure[]{
     \includegraphics[width=7.5cm]{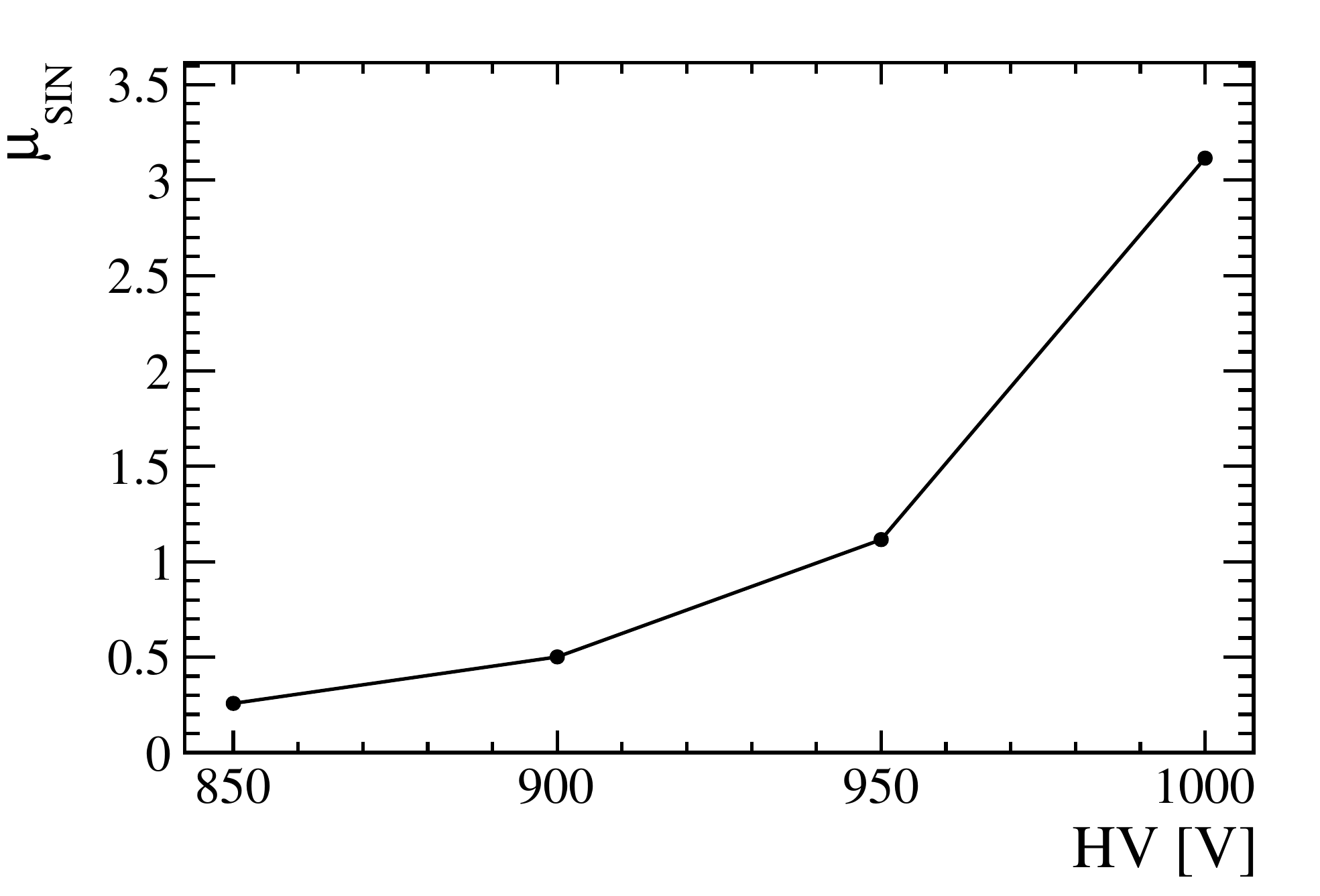}
     \label{HVdependenceRatio}}
   \subfigure[]{
     \includegraphics[width=7.5cm]{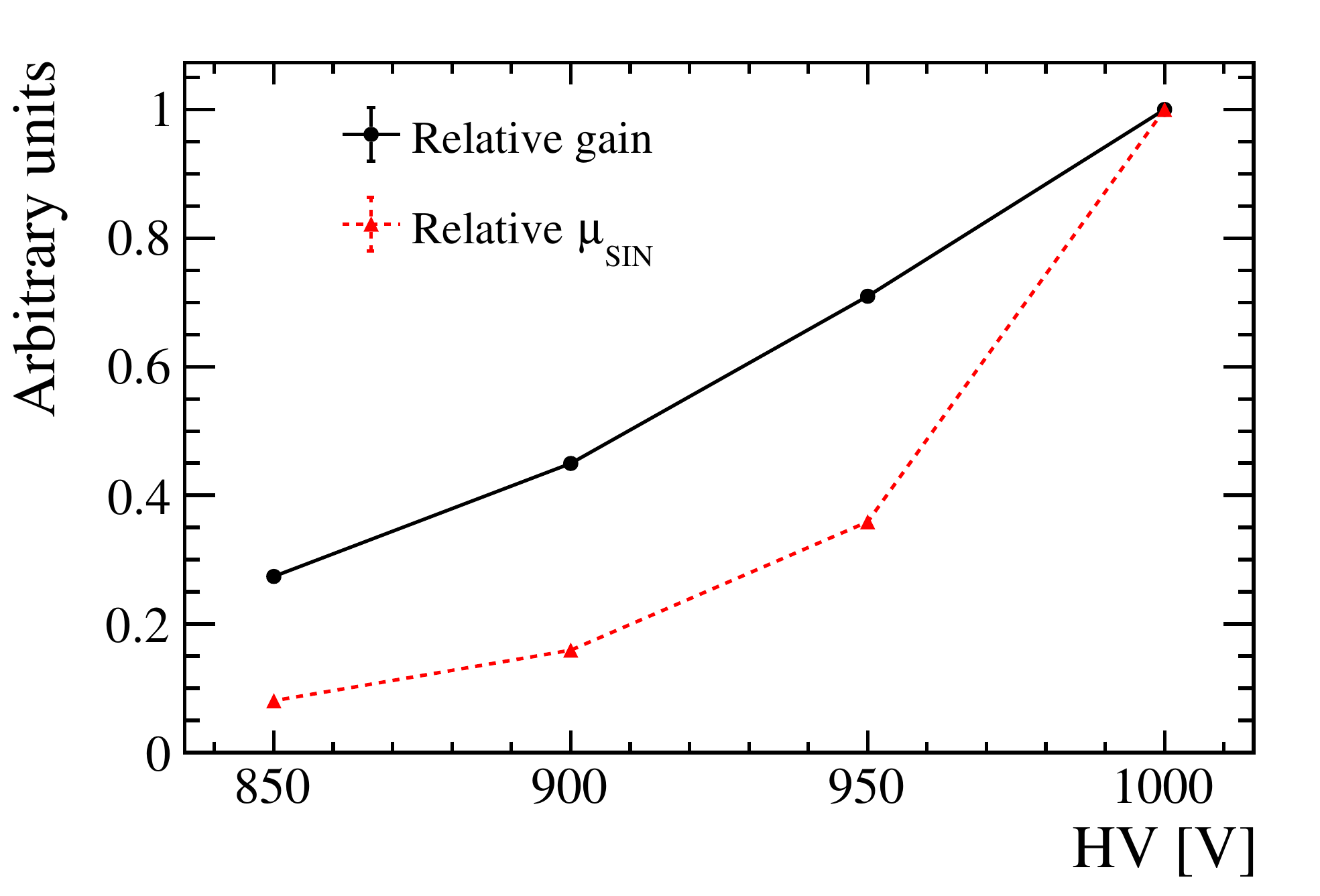}
     \label{HVdependenceRatioAndGain}}
   \caption{\subref{HVdependenceRatio}: mean number of SIN pulses
     as a function of the applied HV for the same
     anode 61 of
     Fig.~\ref{HVdependencePlot}. \subref{HVdependenceRatioAndGain}:
     relative value of gain (black) and of the mean number of SIN
     pulses (red) as a function of the applied HV for the same anode
     and normalised to the measurement at 1000 V.}
  \label{HVdependencePlotsHV}
\end{figure}

The mean number of SIN pulses has an exponential dependence on the high voltage, as shown in Fig.~\ref{HVdependenceRatio}. The comparison between the gain and SIN curves versus the high voltage is reported in Fig.~\ref{HVdependenceRatioAndGain}, where the measurements at 1000 V are used as reference. 

The distributions shown in Fig.~\ref{multipleHVdependence} present similar
peaking structures: the first, second and third peaks appear at $175 \pm
25 \ns$, $300 \pm
25 \ns$ and $400 \pm
25 \ns$ after the signal, respectively, where the 25\ns uncertainty reflects the DAQ time binning. Even if there is no observable shift in time as the high voltage changes, such peaking structures are usually due to ion feedback: in order to identify possible ions contributing to SIN pulses, the transit times can be determined using the classical relation

\begin{equation}
    t = L\sqrt{\frac{2m}{qV}},
    \label{ionTransiTime}
\end{equation}

where $q$ is the ion charge, $m$ is the ion mass, $V$ and $L$ are the potential difference and the distance, respectively, between the start and end point of the ion trajectory, and a uniform electric field is assumed. The voltage divider in the baseboard provides a potential difference between the photocathode and the first dynode of $V_1 = 153.3$ V and 130.3 V when biasing the MaPMT at 1000 V and 850 V, respectively. The distances between the photocathode and the anode, and between the photocathode and the first dynode, are estimated to be 13\mm and 2\mm, respectively. The assumption of a uniform electric field between the photocathode and the first dynode is a very good approximation, besides the corner pixels where small edge effects due to the focusing optics are expected. More significant edge effects affect the configuration of the electric field between the photocathode and the anode, in particular on the peripheral pixels, since the walls of the MaPMT are at the same potential as that of the photocathode. For this reason, Eq.~\ref{ionTransiTime} only gives an approximate value of the transit times between the photocathode and the anode. The transit times calculated from the relation above for various ion candidates~\cite{Haser:2013is,Giordano:2015zqd,KM3NeT:2018nwm}, for 1000 V and 850 V biases, and between photocathode to first dynode and photocathode to anode are reported in Tab.~\ref{tabIonTransitTimes}.

\begin{table}[tb]
  \caption{Transit times, expressed in ns, for various ion candidates reported in the first column. The corresponding atomic masses A are reported in the second column and expressed in units of atomic mass (1 u = $1.66 \times 10^{-27}$ kg). The other columns represent the transit times ($\Delta t$) for, from left to right, photocathode (PK) to first dynode (Dy) at 1000 V, photocathode to first dynode at 850 V, photocathode to anode (An) at 1000 V, photocathode to anode at 850 V, respectively.}
\begin{center}
\begin{tabular}{l c | c c | c c}
    \midrule
    Ion & A [u] & $\Delta t_{\text{PK-Dy}}$ [ns]& $\Delta t_{\text{PK-Dy}}$ [ns]& $\Delta t_{\text{PK-An}}$ [ns] & $\Delta t_{\text{PK-An}}$ [ns]\\
    & & V=153.3 V & V=130.3 V & V=1000 V & V=850 V \\
    \midrule
    $\text{H}_2^+$ & 2.0 & 33.0 & 35.8 & 84.0 & 91.1 \\
    $\text{He}^+$ & 4.0 & 46.5 & 50.5 & 118.4 & 128.4 \\
    $\text{CH}_4^+$ & 16.0 & 93.1 & 101.0 & 237.0 & 257.1 \\
    $\text{K}^+$ & 39.1 & 145.4 & 157.7 & 370.1 & 401.4 \\
    $\text{Sb}^+$ & 121.8 & 256.6 & 278.3 & 653.0 & 708.3 \\
    $\text{Cs}^+$ & 132.9 & 268.1 & 290.7 & 682.3 & 740.0 \\
    \midrule
  \end{tabular}
\end{center}
\label{tabIonTransitTimes}
\end{table}

\begin{figure}[tb]
   \centering
   \subfigure[]{
     \includegraphics[width=7.5cm, keepaspectratio]{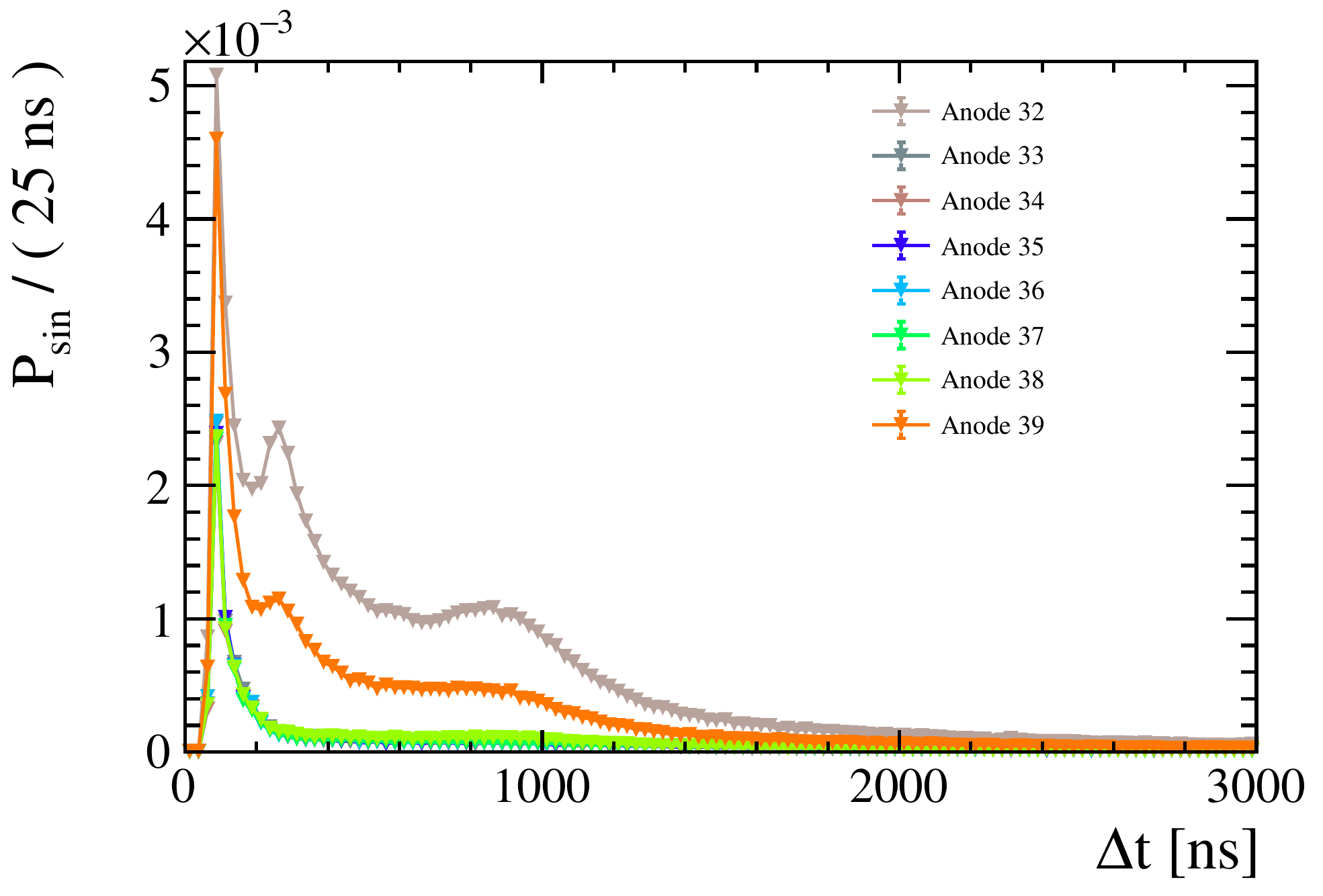}
     \label{notAffectedLines}}
\subfigure[]{
 \includegraphics[width=7.5cm]{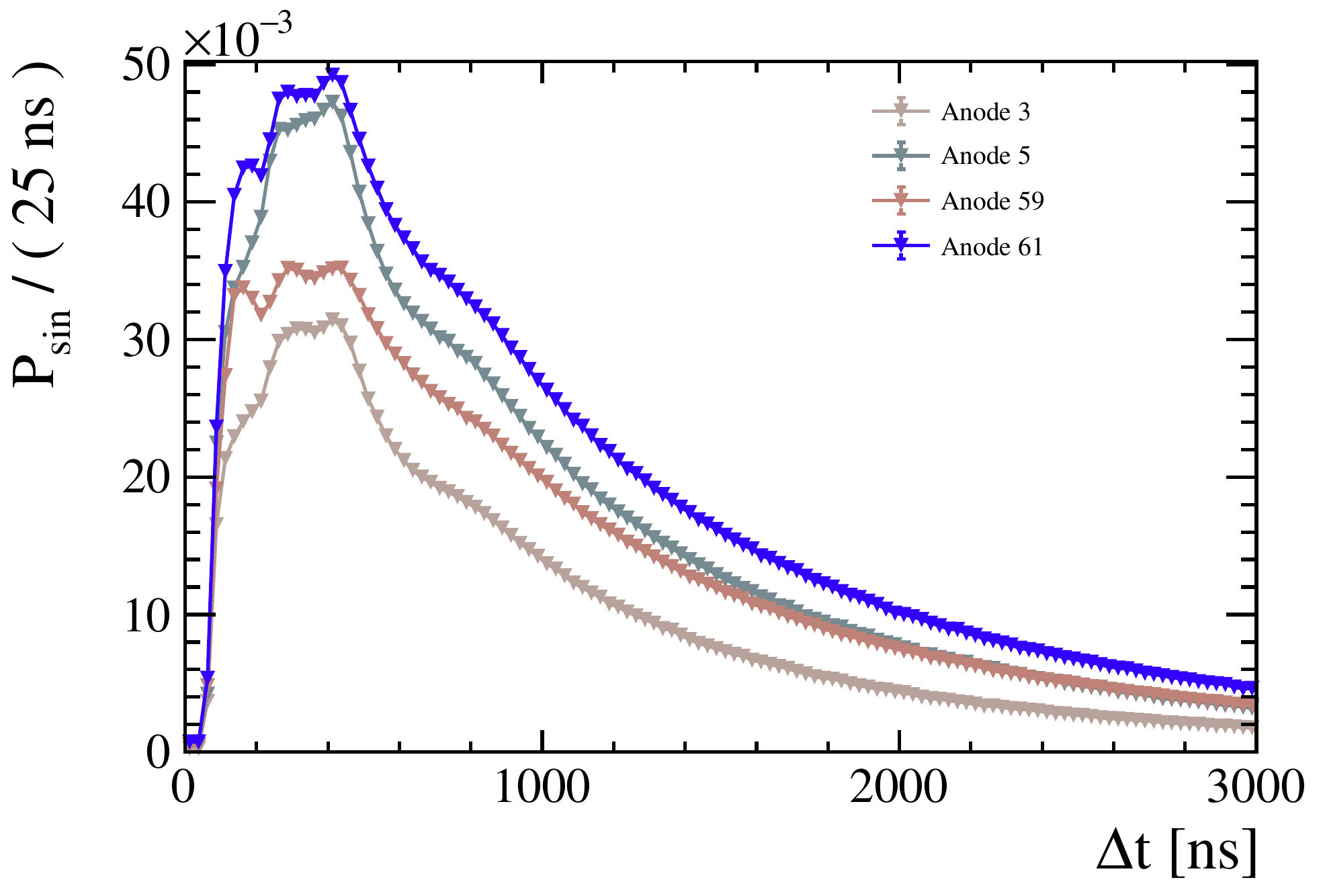}
     \label{affectedLines}}
   \caption{\subref{notAffectedLines}: SIN pulses per-bin probability for the least affected pixels. \subref{affectedLines}: SIN pulses per-bin probability for the most affected pixels. Note that the maximum probability is approximately ten times more with respect to \subref{notAffectedLines}.}
  \label{sinComponents}
\end{figure}

In order to gather information on the nature of SIN, the same per-bin SIN probability, averaged over 300 MaPMTs, is studied for other pixels, as shown in Fig.~\ref{sinComponents}. Different components are present depending on the pixel under consideration. Fig.~\ref{notAffectedLines} refers to a central row of pixels, for which the SIN probability is at least a factor ten lower, on average, with respect to the most affected top and bottom rows. A narrow peak centred at $\Delta t = 100 \ns$ after the signal is visible in all the pixels, while two other peaks, centred at $\Delta t = 275\ns$ and a wider one at $\Delta t = 850\ns$ are only visible for the peripheral pixels 32 and 39, but also present with at least a 20 times lower probability for the neighbouring pixels, with decreasing rate while moving towards the centre of the tube.

  Such peaking structures, and the localisation of SIN events described in Sec.~\ref{sec:localisation}, can be interpreted considering the internal structure of the photon detectors. Mechanical apertures are present at the tube periphery in the top and bottom regions most affected by SIN effects and, to a lesser extent, in the left- and right-most columns. Photons and ions produced at the anode can therefore propagate freely through these apertures towards the photocathode. In the central region of the tube, the dynode chains are screening these effects. The narrow peak centred at $\Delta t = 100 \ns$ can be associated to $\text{CH}_4^+$ feed-back between the photocathode and the first dynode. The other structures at $\Delta t = 275\ns$ and $\Delta t = 850\ns$ are compatible with $\text{CH}_4^+$ and $\text{Cs}^+$ feed-back, respectively, between the photocathode and the anode. A component corresponding to $\text{Cs}^+$ feed-back between the photocathode and the first dynode could be present but shadowed by the tail of the first peak. Fig.~\ref{affectedLines} displays the SIN time distribution for four pixels belonging to the most affected rows. The dominant feature of such distributions is the exponential tail, indicating the presence of a mechanism consistent with internal light emission and fluorescence decay. The positions of the peaking structures are shifted towards later times due to the convolution with such effects and to the actual electric field configuration, and are compatible with both ion feed-back and cascade processes. The time gap of approximately $200 \ns$ between the signal and the first peak suggests that the internal light emission is mainly amplified by the larger number of photoelectrons produced by the interaction between ions and the photocathode. A shoulder around 800\ns is also visible and can be associated to the same effects observed in the central row of pixels due to $\text{Cs}^+$ feedback between the photocathode and the anode. Details on electron-impact ionisation cross-sections for $\text{CH}_4$ and $\text{Cs}$ can be found in Refs.~\cite{doi:10.1063/1.473468} and \cite{PhysRevA.74.032708}, respectively.

\subsection{Analogue measurements}
\label{sec:sinAnalogue}

\begin{figure}[tb]
   \centering
     \includegraphics[width=10cm, keepaspectratio]{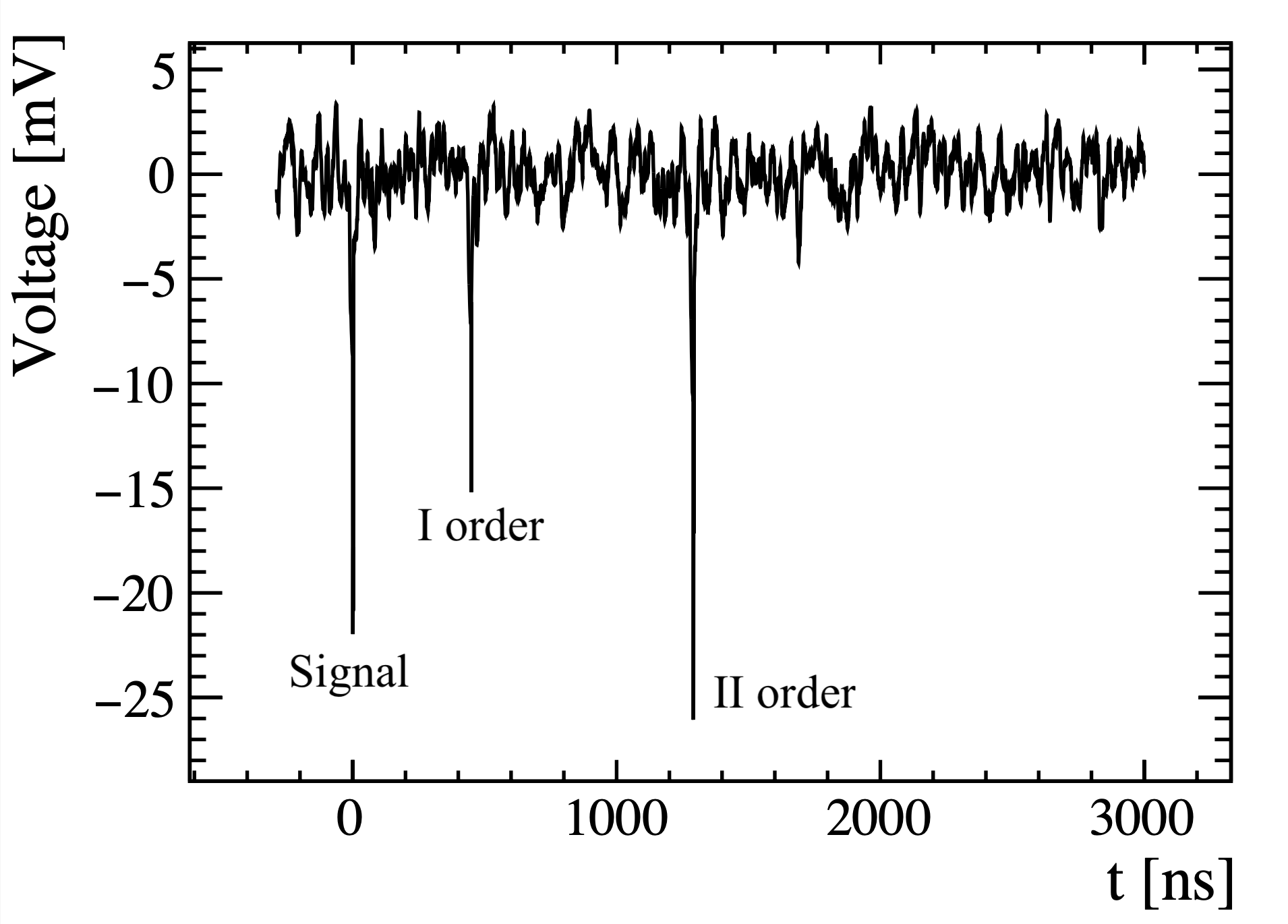}
   \caption{Example of a waveform recorded with the oscilloscope. The MaPMT response to the laser signal is in correspondence of t=0\ns, while two SIN pulses can be observed at later times.}
  \label{scopeWaveform}
\end{figure}

Further investigations have been performed with an oscilloscope by probing the output of one MaPMT channel, in order to validate the interpretation of the time distribution of SIN events. An example of recorded waveform showing the presence of out-of-time events is reproduced in Fig.~\ref{scopeWaveform}.

\begin{figure}[tb]
   \centering
     \includegraphics[width=15cm, keepaspectratio]{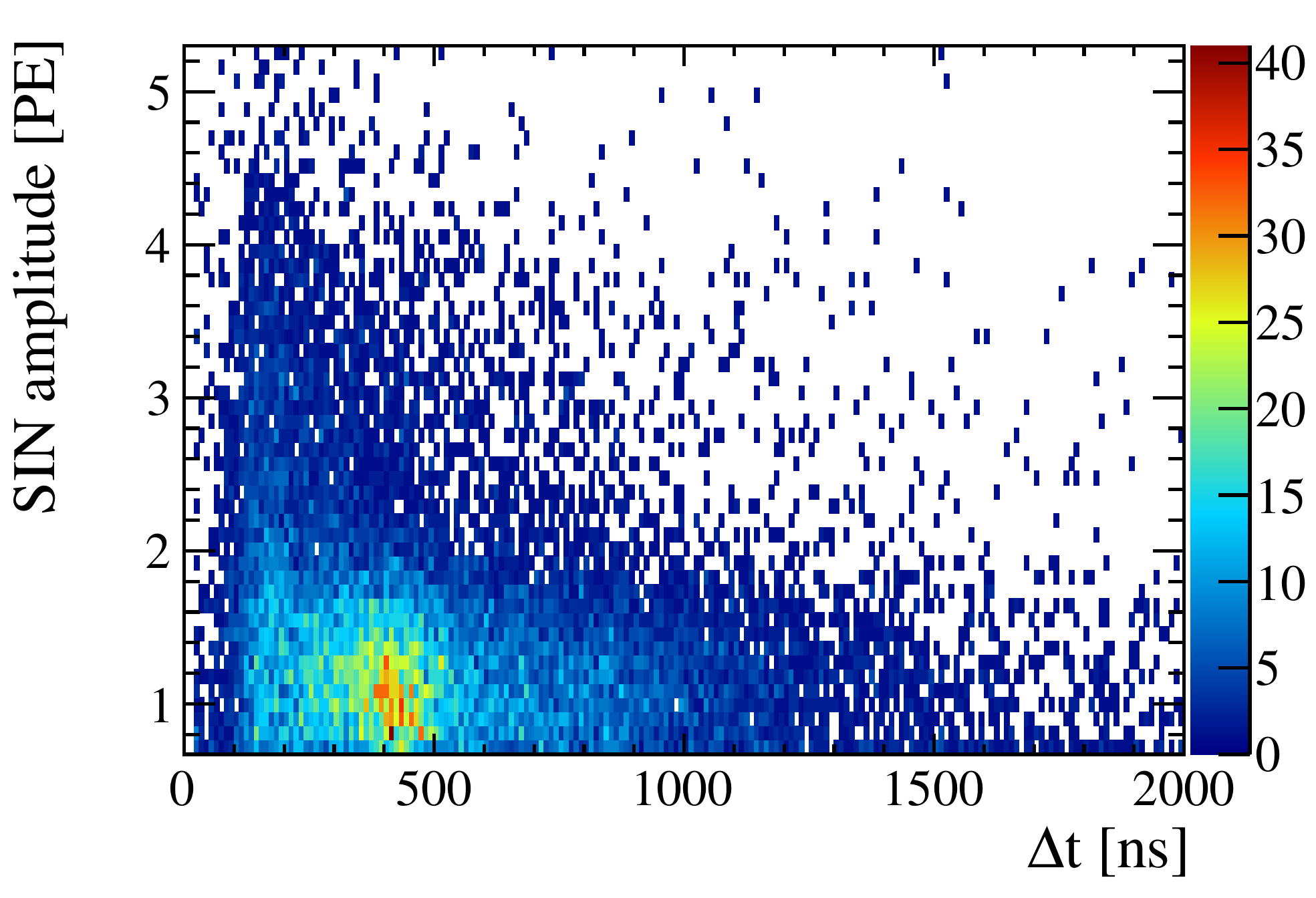}
   \caption{SIN amplitudes expressed in single-photoelectron (PE) units as a function of the elapsed time from the signal.}
  \label{oscAmp}
\end{figure}

The data collected with the oscilloscope, though statistically limited, are used to draw qualitative conclusions on the nature of SIN pulses. In particular, the SIN amplitude versus the elapsed time is determined as reported in Fig.~\ref{oscAmp}. The main contribution to SIN events, corresponding to the region around 400\ns is due to pulses having single-photon amplitude. In this region, the dominant mechanism responsible for this source of noise is attributed to light emission generated from the anode side of the MaPMTs. 

The events in the band centred at $\Delta t = 200\ns$ have larger amplitudes, equivalent to up to four photoelectrons. This cluster of events confirms the interpretation of the first peak as due to ion feed-back. However, the rate of ion feed-back events is approximately one order of magnitude lower with respect to the SIN pulses produced by the dominant light emission mechanism.

\begin{figure}[tb]
   \centering
     \includegraphics[width=15cm, keepaspectratio]{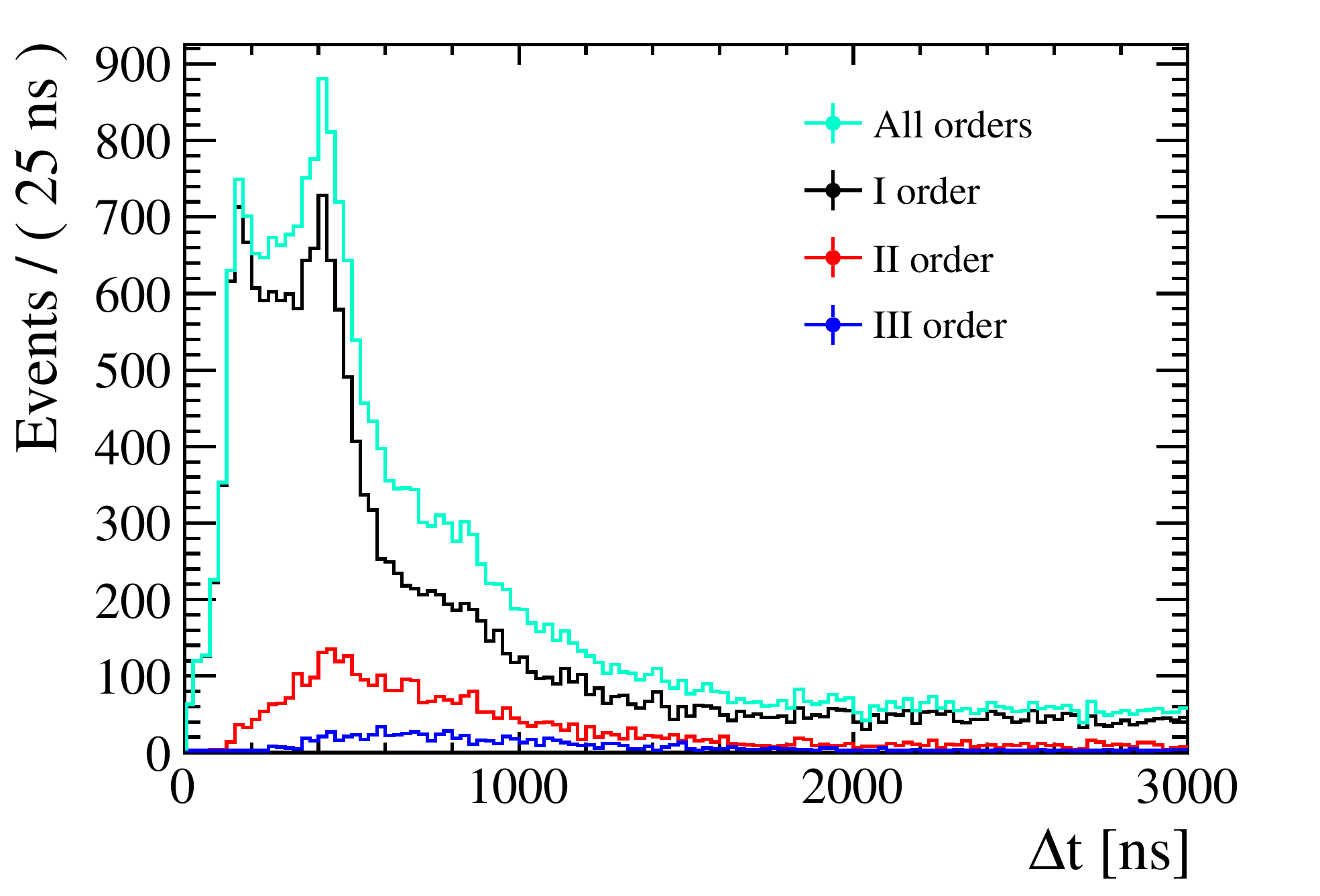}
   \caption{SIN time distribution with order components as described in the legend.}
  \label{orders}
\end{figure}

The waveforms acquired with the oscilloscope can also be used to determine the amount of higher-order events, \ie noise signals generated by photoelectrons arising from SIN pulses rather than from primary signal photons, for the corresponding pixel of this particular MaPMT. Such higher-order components are shown in Fig.~\ref{orders} in which the majority of events are seen to be of the first order, \ie the peaking structures are mainly due to independent mechanisms and not to cascade effects. Approximately 16\% of the total is due to second-order SIN pulses that accumulate around $\Delta t = 400\ns$, while around 4\% are due to the third order. However, the amount of higher order SIN pulses may be underestimated since the primary may be present in adjacent pixels and not detected in this particular anode.

\section{Pile-up studies}
\label{sec:pileUp}

As described in Sec.~\ref{sec:analysis}, the SIN time distribution extends over several microseconds after the primary signal. The probability to detect noise events is conditioned to the presence of signal in previous time slots. As a consequence, in case of a high illumination rate, the
detected photon occupancy is increased in the regions affected by SIN effects. As determined from simulations, the spatial distribution of Cherenkov photons across the photon detector planes of the LHCb RICH detectors is expected to vary
significantly, with a peak detection rate of $10 \mhz$ in
the central region of \richone, corresponding to an occupancy of approximately 30\%.

The pile-up studies were performed in the laboratory at \cern, employing an upgrade photon detector module, a pcie40-based DAQ and slow control and a laser-based illumination system. The acquisition is synchronous with the laser pulses, the trigger signal being sent to both the DAQ boards and the laser driver. The DAQ time window is one 25\ns slot wide and a trigger rate $\nu_t$ is equivalent to introducing a dead time $\tau_d = 1/\nu_t - 25 \ns$ in the DAQ system. Measurements for $\nu_t$ up to $10 \mhz$ are performed in order to
evaluate the increase in detected counts due to SIN pulses. The laser intensity is tuned to get an occupancy of approximately $1\%$ when running at $\nu_t =
100 \khz$, corresponding to a dead time $\tau_d = 10 \mus$, \ie larger than the typical duration of SIN pulses. 

The probability $\bar{P}_m$ to have no hits detected by a pixel in a
25\ns time slot $m$ is given by the probability $\bar{P} =
e^{-(\mu_s+\mu_d)}$ to detect no signal or dark counts in this time
slot multiplied by the probability to have no SIN pulses in time slot $m$ due to
signal or dark counts in previous time slots. $P_{\text{ck}} =
1-\bar{P_{\text{ck}}}$ indicates the detection probability without 
SIN, with the subscript ck to indicate Cherenkov signal events (or any signal other than SIN background
sources), that can be determined from simulation and takes the same mean value for each 25\ns time slot. The SIN pulse probability can be determined from the measurement of the SIN time distribution described in Sec.~\ref{sec:timeDistribution}. Since this distribution includes the contribution from higher-order SIN pulses, the probability to have hits in slot $m$ is given by 

\begin{align}
P_{m}  = 1-\bar{P}_{\text{ck}} \prod_{i=1}^m(1-P_{\text{sin},i}P_{\text{ck}}),
\label{probNoCounts}
\end{align} 

where $P_{\text{sin},i}$ denotes the probability to detect a
SIN pulse $i$ slots after the corresponding detected photon or dark count. The duration of SIN pulses given a signal is limited in time and their probability is null $M$ time slots after the signal, where $M$ corresponds to the maximum duration of SIN pulses (\eg $M=160$ for a duration of 4\mus). Consequently, the probability to detect hits in the presence of SIN for $m>M$ converges to

\begin{align}
P  = 1-\bar{P}_{\text{ck}} \prod_{i=1}^M(1-P_{\text{sin},i}P_{\text{ck}}) \approx
  1-\bar{P}_{\text{ck}} e^{-\sum_{i=1}^MP_{\text{sin},i}P_{\text{ck}}} = 1-\bar{P}_{\text{ck}} e^{-\mu_{\text{sin}}P_{\text{ck}}},
\label{probNoCountsFinal}
\end{align} 

where the approximation $P_{\text{sin},i}P_{\text{ck}} \ll 1$ is used and
$\mu_{\text{sin}}$ is the mean number of SIN pulses. The
probability is now independent from the slot $m$ under study.

The introduction of a dead time, resulting from the DAQ strategy in
the high-rate measurements described in this section, requires the
rewriting of Eq.~\ref{probNoCounts} as

\begin{align}
P_{m}  = 1-\sum_{n=0}^N \delta (m - n \Delta m)  \left [ \bar{P}_{\text{ck}}
  \prod_{i=1}^m \left ( 1-\sum_{k=1}^{m/\Delta m} \delta (i - k \Delta
  m) P_{\text{sin},i}P_{\text{ck}} \right ) \right ],
\label{probNoCountsDeadTime}
\end{align} 

where $N=10^{6}$ is the total number of acquired time slots, \ie the number
of trigger pulses, and $\Delta m$ is
the number of $25 \ns$ time slots between each trigger pulses, given
by $\Delta m = 40 \mhz/\nu_t$. When $\nu_t = 100 \khz$, the probability
to detect a hit in the acquired slot $m$ is equal to the counting
probability without SIN, $P_{m}  =
1-e^{-(\mu_s+\mu_d)}$, since $\Delta m = 400$ and
$P_{\text{sin},k\Delta m}=0$. No dead time, \ie when the acquisition
runs at $40 \mhz$ as expected during the data taking corresponding to the collisions
mode at the LHC, means $\Delta m=1$ and Eq.~\ref{probNoCountsDeadTime}
reduces to Eq.~\ref{probNoCounts}.

\begin{figure}[tb]
   \centering
\subfigure[]{
 \includegraphics[width=7.5cm]{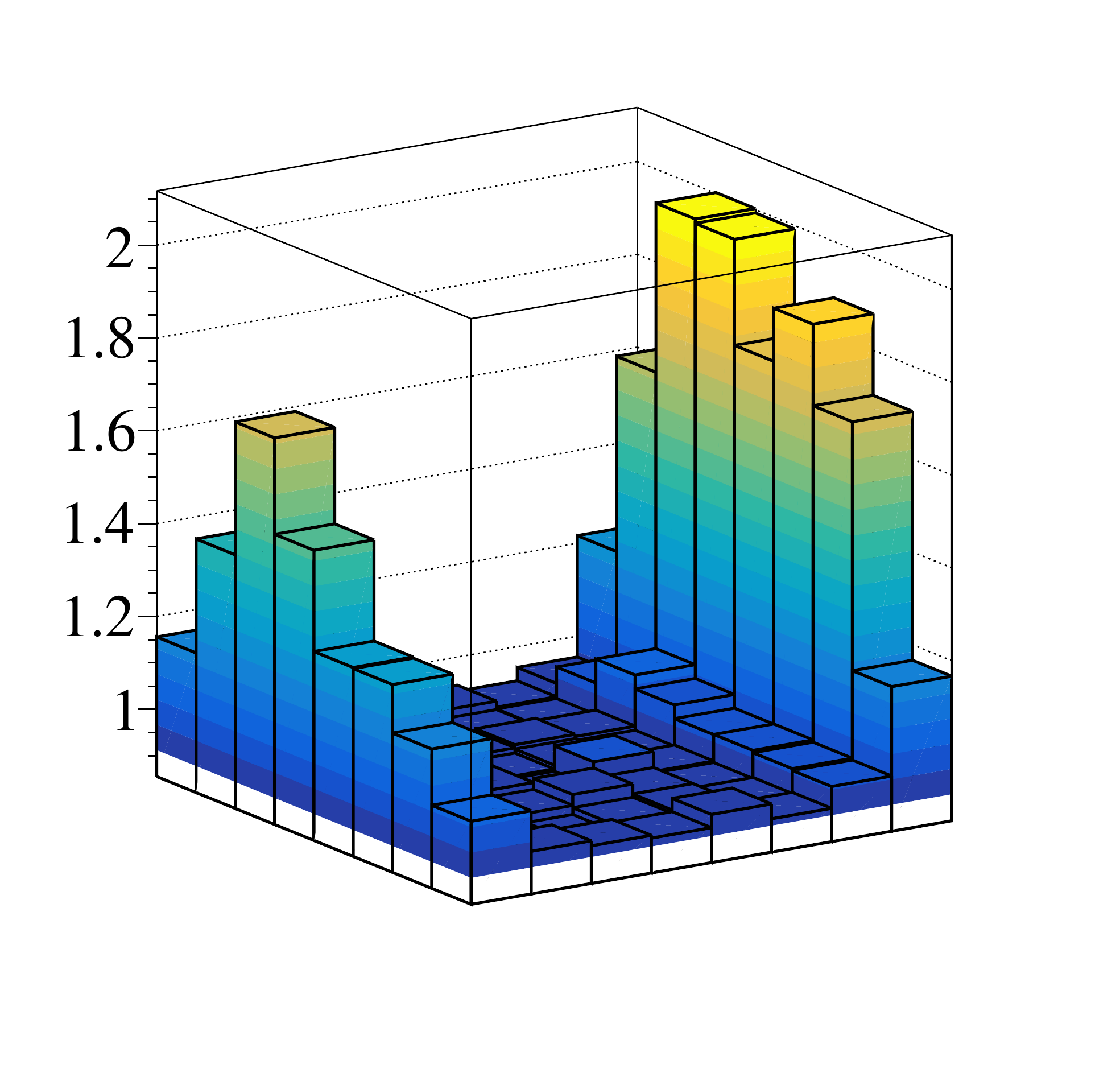}
     \label{pathological_rot}}
   \subfigure[]{
     \includegraphics[width=7.5cm, keepaspectratio]{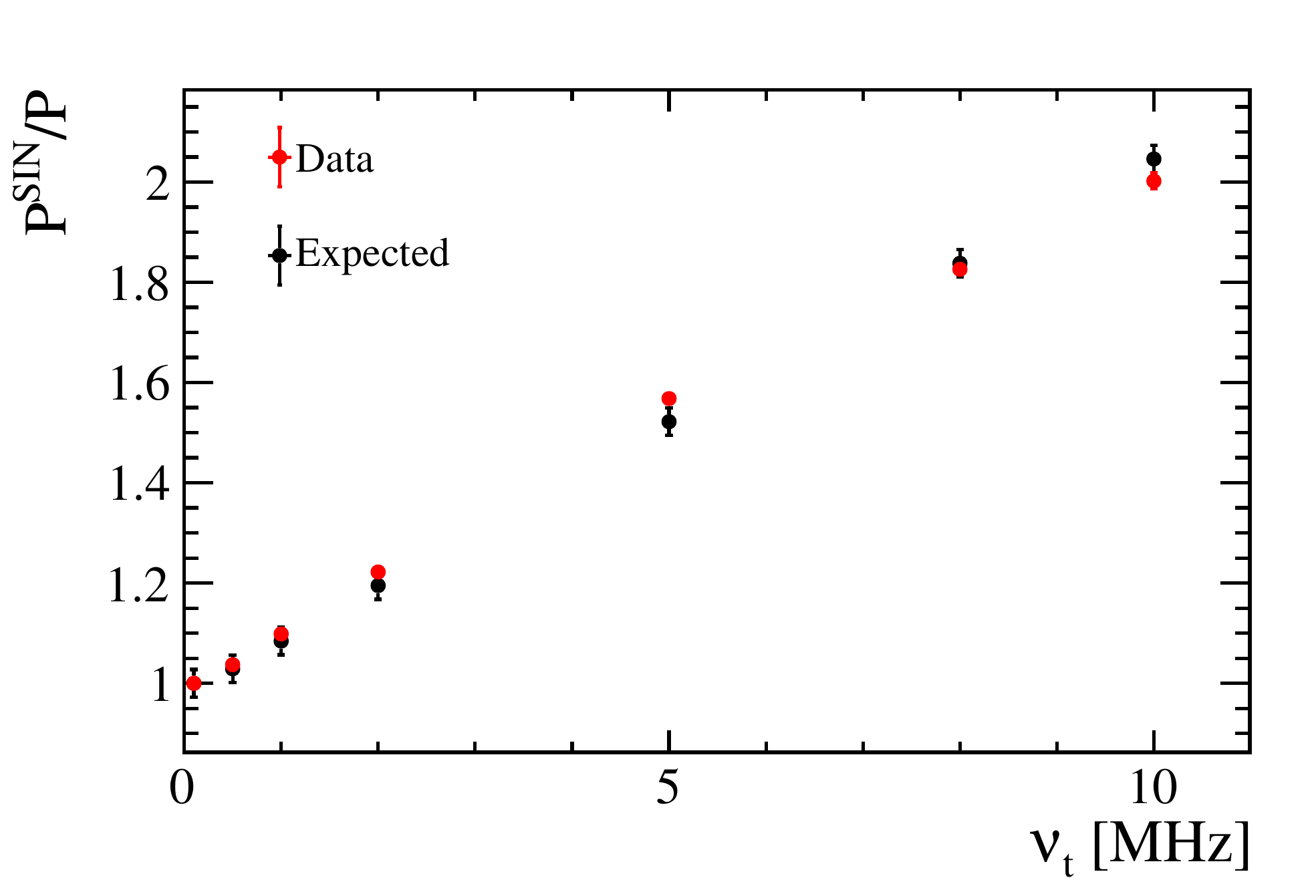}
     \label{pathologicalModel}}
   \caption{\subref{pathological_rot}: ratio of detected counts at $10 \mhz$ and $100 \khz$ for MaPMT FB3036. \subref{pathologicalModel}: ratio of the counting probabilities with and without SIN for pixel 61 of MaPMT FB3036, at different illumination rates, for the
     experimental (red) and expected (black) values. The mean number of SIN pulses is $\mu_{\text{SIN}}^{61} = 4.329 \pm 0.031$.}
  \label{occupancyRatio}
\end{figure}

To check the above relations, the ratio between the hit maps acquired at different illumination rates is determined. The ratio between the occupancy at 10\mhz and 100\khz for one MaPMT is shown in Fig.~\ref{pathological_rot}. The expected values based on the model described above agree with the measurements, as reported in Fig.~\ref{pathologicalModel} for one pixel. In particular, the validity of Eq.~\ref{probNoCountsFinal} allows to assess the impact of SIN on the PID performance, described in Sec.~\ref{sec:PID}, with the knowledge of the mean number of SIN pulses and the occupancy values across the \lhcb \rich photon detectors planes. Other measurements at larger occupancies, \eg at 10\%, have been performed and give consistent results.

\section{Mitigation strategies}
\label{sec:mitigation}

Different strategies to mitigate the SIN effects are described in the present section. They will be adopted to operate the R11265 MaPMTs in the 40\mhz single-photon counting mode.

\subsection{Optimisation of the high-voltage operating point}

Given the strong dependence of the mean number of SIN pulses on the applied HV, the MaPMTs will be operated at the lowest possible HV, as a compromise between a high single-photon detection efficiency and a low SIN background rate. While operations at low HV is sufficient to recover the expected performance for a large region of the \rich detectors acceptance, additional strategies are required in the central region of \richone where an average illumination rate of 10\mhz is expected.

\subsection{New MaPMTs with reduced SIN effects}

\begin{figure}[tb]
   \centering
\subfigure[]{
 \includegraphics[width=7.5cm]{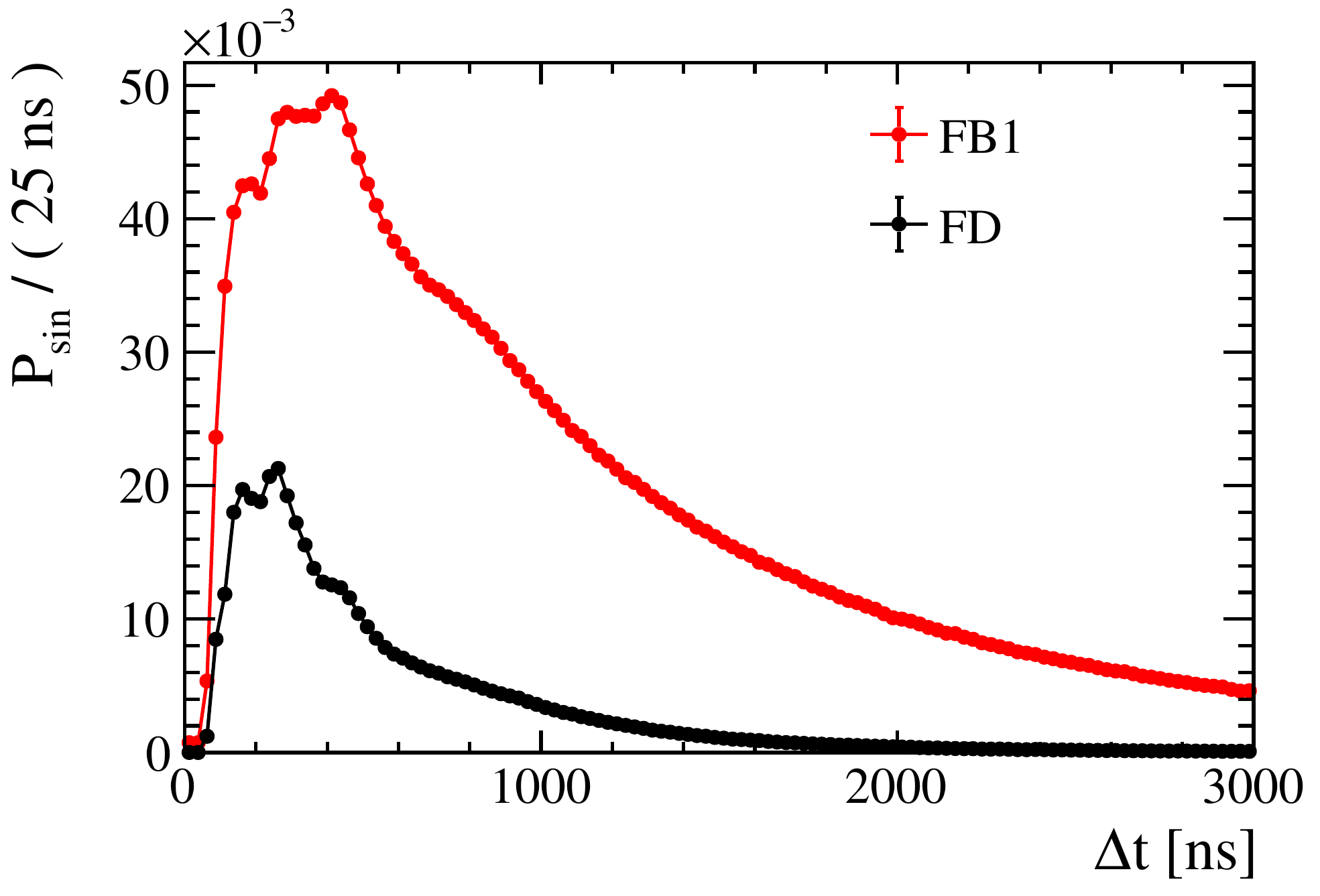}
     \label{FDFBcomparison}}
   \subfigure[]{
     \includegraphics[width=7.5cm, keepaspectratio]{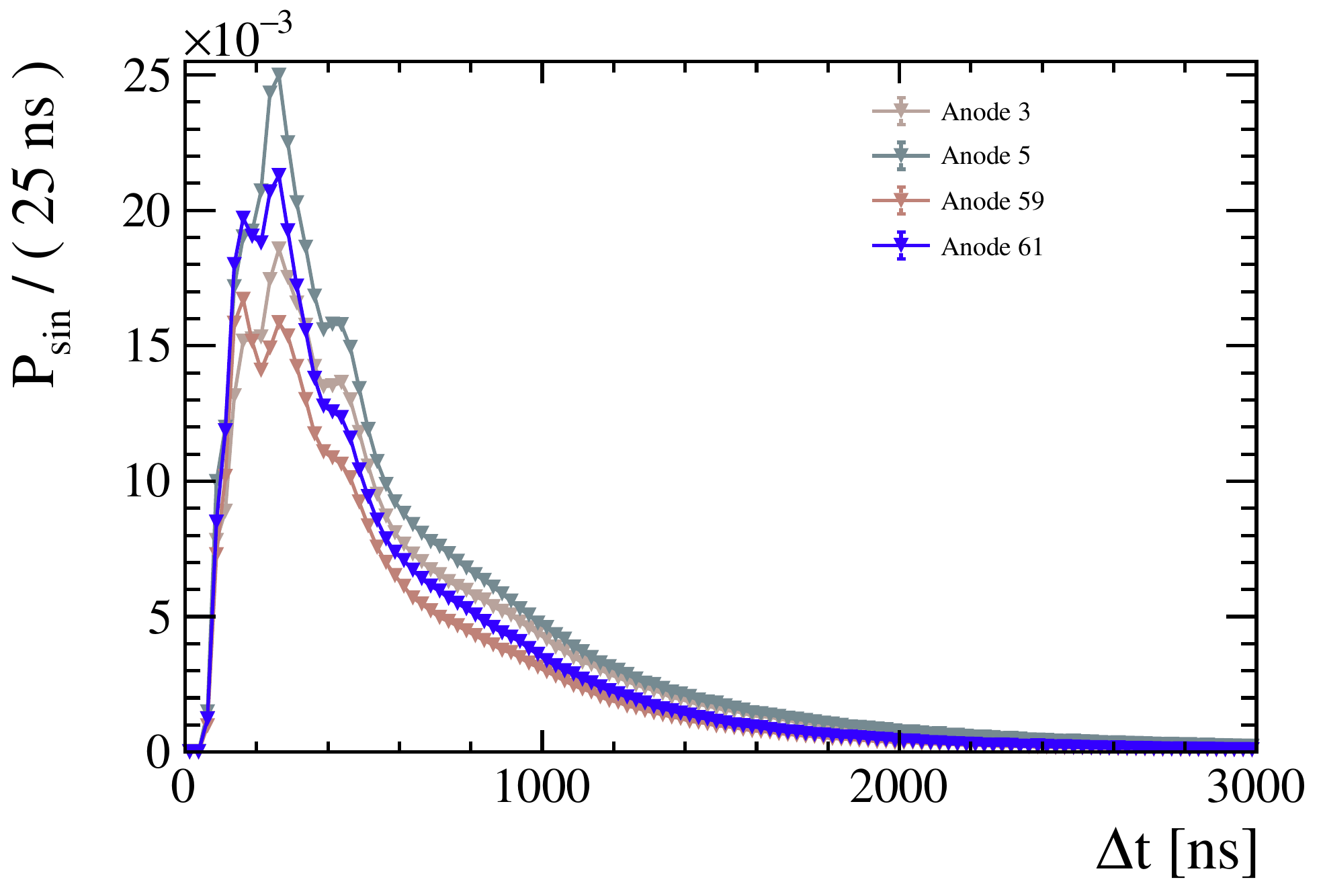}
     \label{FDtimeStructures}}
   \caption{\subref{FDFBcomparison}: Comparison of SIN distributions between the original FB-series and the new FD-series MaPMTs. The maximum per-bin probability is more than halved, with a faster recovery of tail events. \subref{FDtimeStructures}: per-bin SIN pulse probability for 4 pixels of the FD series.}
  \label{hamaMitigation}
\end{figure}

A new series of R11265 MaPMTs, referred to as the FD series, has been produced by Hamamatsu to reduce the contributions from SIN pulses. The comparison of the per-bin SIN pulse probability of pixel 61 between the FB and FD series, averaged over 200 MaPMTs, is reported in Fig.~\ref{hamaMitigation}. A change in the internal mechanical design of the tube results in a large reduction of SIN effects. The FD-series MaPMTs will be used in the high-occupancy region of \richone.

\subsection{Nanosecond time gate at the front-end readout}
\label{sec:mitigation:timeGate}

\begin{figure}[tb]
    \centering
    \includegraphics[width=0.5\linewidth]{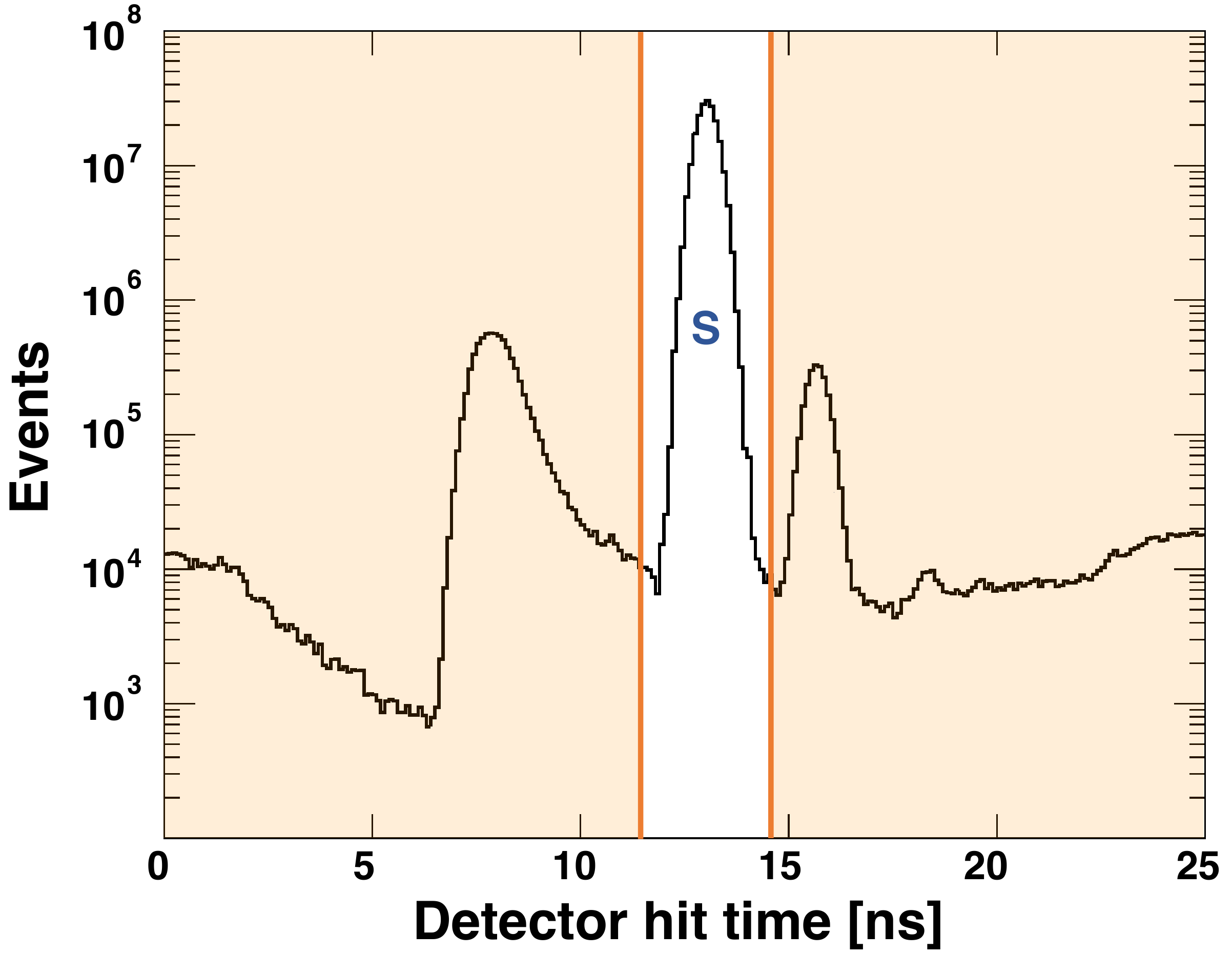}
    \caption{\richone MaPMT photon detector hit time distribution for LHC beam interactions, showing the signal peak (S) within a 3.125\,ns time gate. As indicated by the shaded regions, the time gate excludes MaPMT hits during most of the 25\,ns period between bunch crossings. }
    \label{fig:RICH1timing}
\end{figure}

The prompt emission of Cherenkov radiation and the focusing mirror geometry result in an excellent intrinsic time resolution of the \rich detectors. Per-track Cherenkov photons from a single proton-proton interaction reach the MaPMT plane with a spread of $\mathcal{O}(10\,$ps) for \richone and $\mathcal{O}(100\,$ps) for \richtwo. The larger spread for \richtwo is due to the size and position of the detector. Within the LHC environment the time spread of the signal is dominated by the space and time distributions of the proton collisions at the interaction point. Fig.~\ref{fig:RICH1timing} shows the photon arrival time at the \richone MaPMT plane following a bunch crossing at time zero, simulated in the LHCb framework. A time gate of a few nanoseconds can be used to collect all Cherenkov photon signals (populating the peak labelled S) whilst rejecting background from beam interactions and MaPMT noise. The time gate covers a small fraction of the 25\,ns between LHC bunch crossings and therefore allows up to an order of magnitude suppression of out-of-time background hits in the front-end output data.

During LHC Run~3, the time gate will be implemented in the FPGAs of the PDMDB. The programmable logic in the FPGA is adapted to sample the digital signals from the CLARO front-end chips at 320\,MHz using the de-serialiser embedded in every input-output logic block. The FPGA logic can be configured to detect specific input patterns and to apply a time gate with a minimal width of 3.125\ns around the signals from the CLAROs. The gate can be synchronised with the signal peak by shifting the sampling clock phase with respect to the LHC bunch-crossing clock. Given the overall time resolution of the electronic readout chain, a 6.25\ns time gate may be required to achieve the best PID performance. Since the SIN probability is approximately constant over the relatively short 25\ns time between bunch crossings, the signal-to-SIN ratio is expected to improve by a factor of four or eight using a time gate of 6.25\ns or 3.125\ns, respectively.

\section{Effects on the PID performance}
\label{sec:PID}

\newcommand{\mapmt}{\mbox{MaPMT}\xspace}
\newcommand{\mapmts}{\mbox{MaPMTs}\xspace}
\newcommand{\mapmtFull}{\mbox{Multianode} \mbox{Photomultiplier} Tube\xspace}
\newcommand{\rTypeMapmt}{\mbox{R11265}\xspace}
\newcommand{\hTypeMapmt}{\mbox{R12699}\xspace}
\newcommand{\claro}{\mbox{CLARO}\xspace}
\newcommand{\pp}{\mbox{$pp$}\xspace}
\newcommand{\pid}{\mbox{PID}\xspace}
\newcommand{\mySin}{\mbox{SIN}\xspace}
\newcommand{\run}[1]{\mbox{Run#1}\xspace}
\newcommand{\minBias}{\mbox{minimum-bias}\xspace}

\newcommand{\sinNrOfPulses}{\ensuremath{\mu_{\text{sin}}}\xspace}
\newcommand{\occNoSin}{\ensuremath{P_{ck}}\xspace}
\newcommand{\sinProbPerPixel}{\ensuremath{x_{SIN}}\xspace}

The~\lhcb simulation framework is used to evaluate the~effects of \mySin on the~\pid performance. The~\mySin modelling is introduced during the~digitisation of the~detector response, which is conducted within the~\boole application.

\subsection{Simulation input}
\label{sec:PID:input}

In the~\run{ 3} simulation, various pixel properties, such as the average pixel gain and the mean number of \mySin pulses, will be read from a~dedicated database populated with data from the~commissioning phase and the~quality assurance measurements. Although the~detector installation is ongoing at the~time of writing this paper, the~\mapmts in the~simulation are placed at their expected positions, which allows to reproduce realistic detector properties and assess the~effect of \mySin on the~\pid performance. Due to the~presence of the~\mbox{high-occupancy} areas in the~\richone detector, a~dedicated procedure is applied to group the~\mapmts into ECs based on their gain and \mySin characteristics and then populate them optimally in different occupancy regions.

\subsection{\mySin model implementation}
\label{sec:PID:implementation}

The~implementation of the \mySin model in the~simulation is based on the~considerations described in Sec.~\ref{sec:pileUp}, which take into account the~\mbox{pile-up} effects due to the~expected high \mbox{detection-rates}. As mentioned there, modelling \mySin requires knowledge of \sinNrOfPulses in the~given pixel as well as pixel occupancy values in the~absence of \mySin (corresponding to \occNoSin). The~latter are based on the~simulation and have the~same average values throughout all $25 \ns$ time slots.

Additional contributions due to \mySin are added in each pixel with a~probability \sinProbPerPixel that satisfies the~relation $P = \occNoSin + \sinProbPerPixel$ (both contributions to the~occupancy are treated independently here), where $P$ is the~detection probability in a~$25 \ns$ \mbox{time-slot} including the~presence of \mySin, as defined in Sec.~\ref{sec:pileUp}. Following Eq.~\ref{probNoCountsFinal} the~\mySin deposits are added to each event at the~digitisation stage with the~probability:
\begin{align}
    \sinProbPerPixel = \left( 1 - \occNoSin \right) \left( 1 - e^{ -\sinNrOfPulses{} \occNoSin }  \right) ,
    \label{sinProbabilityPerPixel}
\end{align} 
which depends on the~pixel properties and its position on the~photon detector plane, as indicated before.

Since a~full \mbox{time-simulation} of the~\rich detector response is performed, \mySin contributions are also added in the~two $25 \ns$ \mbox{time windows} preceding the~signal one, and a~random \mbox{time-of-arrival} is assigned to them within each of the~slots. It allows to emulate a~situation in which a~SIN pulse arrives before the~signal, which can result in the~failure to record an~actual Cherenkov hit. Contributions of all types (meaning the~signal ones, the~\mySin ones, \etc) are then processed in the~same way to provide the~final \claro time response. In the~current simulation, this response is sampled into eight $3.125 \ns$ slots and the~patterns accepted for further processing can be configured (which aims to reproduce the~behaviour of the~\rich digital boards). In a~nominal case, a~pattern configuration corresponding to a~level detection in a~specified $3.125 \ns$ slot after the~signal peak is applied, but more refined scenarios with a~time gate around \claro signals, as described in Sec.~\ref{sec:mitigation:timeGate}, are explored as well.

\subsection{Simulation results}
\label{sec:PID:results}

\begin{figure}[tb]
    \centering
    \subfigure[]{
        \includegraphics[height=6.15cm]{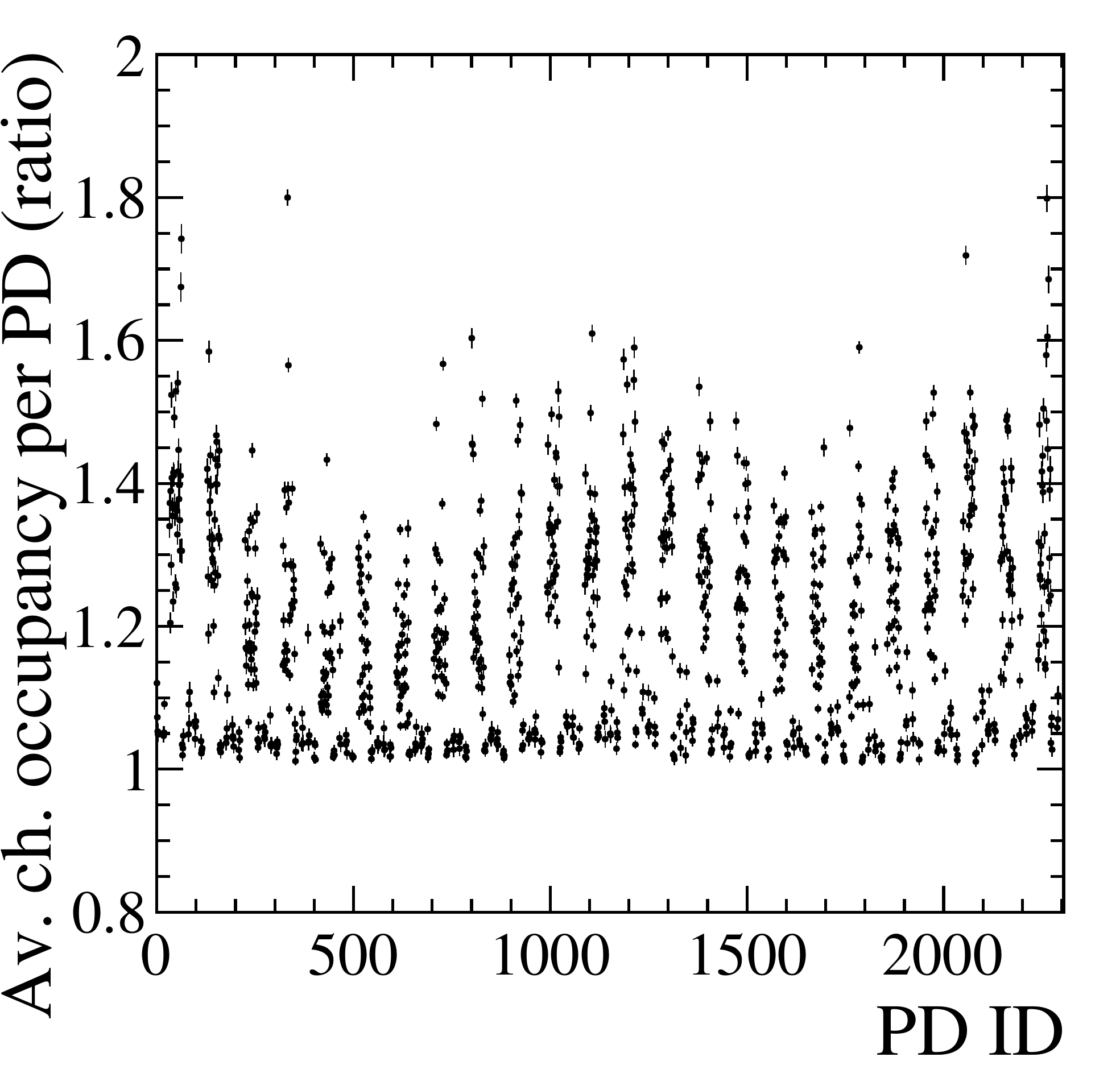}
        \label{fig:occupancyRatio}
    }
    \subfigure[]{
        \includegraphics[height=6.15cm, keepaspectratio]{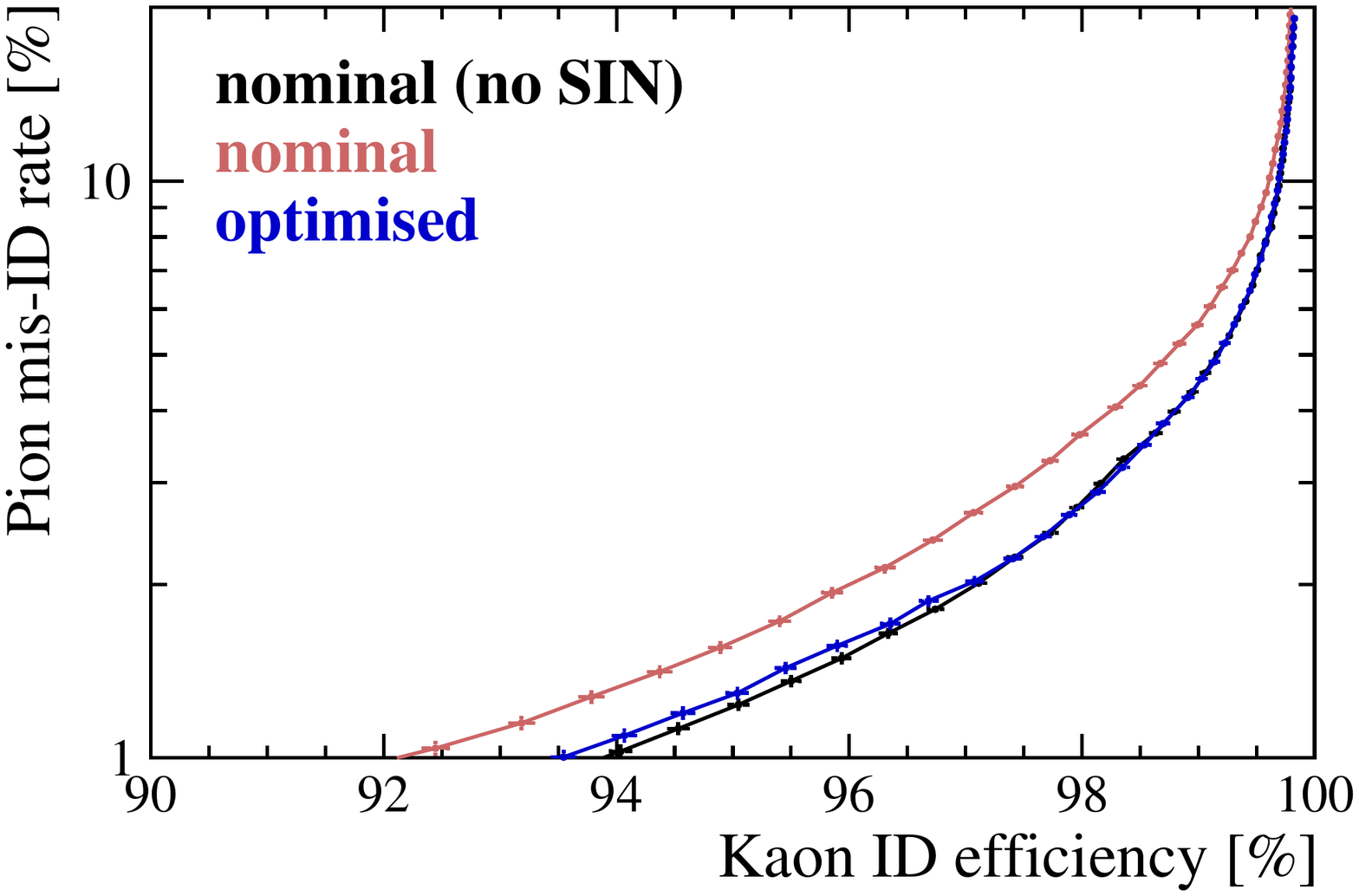}
        \label{fig:PID}
    }
    \caption{
        \subref{fig:occupancyRatio}: Ratio of the~average channel~(pixel) occupancy per \mapmt in \richtwo from the~\lhcb simulation with and without \mySin using the~nominal detector configuration described in the~text. \subref{fig:PID}:~Comparison of the~\pid performance obtained from the~\lhcb simulation including \mySin with the~nominal~(red) and optimised~(blue) scenarios as stated in the~text. A~case corresponding to the~nominal configuration without simulating the~\mySin effect~(black) is also shown for reference. The~curves illustrate the~probability of misidentifying a~pion as a~kaon, as a~function of the~efficiency of identifying a kaon.
    }
    \label{fig:simulation}
\end{figure}

The~effects of the~presence of \mySin, as evaluated using the~simulation, are presented in Fig.~\ref{fig:simulation}. Two scenarios of the~detector configuration are considered.  The~nominal case corresponds to the~pixel properties while operating the~\mapmts at $1000 \unit{V}$ with the~standard \mbox{level detection} described in Sec.~\ref{sec:PID:implementation}. An~optimised one implements some of the~mitigation strategies described in Sec.~\ref{sec:mitigation}. It refers to using the~operating point of $900 \unit{V}$ with the~\mapmts in \richone populated according to the~procedure mentioned in Sec.~\ref{sec:PID:input} and a~preliminary configuration with an~edge detection and a~$9.375 \ns$ \mbox{time gate} around the~signal peak~(defined separately for \richone and \richtwo). Both scenarios account for \mySin, as well as spillover effects from adjacent bunch crossings.

A~substantial increase in the~average pixel occupancy per~\mapmt due to the~additional \mySin deposits is observed in the~nominal scenario~(see Fig.~\ref{fig:occupancyRatio}). The~points at values close to unity correspond to the~\hTypeMapmt \mapmts, where no significant \mySin effect is found. Introducing the~\mbox{\mySin-mitigation} strategies in the~optimised configuration allows to considerably reduce the~increase in the~occupancy related to \mySin to a~level of $< 10\%$ in all \mapmts, which is crucial in terms of an~optimal operation of the~\rich detectors. It is worth noting that despite the~essential contributions from \mySin deposits in the~nominal scenario, the~\pid algorithms are relatively robust with respect to the~sources of random noise like \mySin. Nevertheless, a~sizeable improvement in the~\pid performance is observed after implementing the~\mbox{\mySin-reduction} measures, as illustrated in Fig.~\ref{fig:PID}, and, in this context, the~effect of \mySin appears to be nearly fully alleviated~(see the~reference curve for the~nominal scenario without simulating \mySin). Preliminary studies suggest that, after introducing the~mitigation strategies described in Sec.~\ref{sec:mitigation}, \mySin should have no significant impact on physics performance during the~\run{ 3} operations. 

\section{Conclusions}
\label{sec:conclusions}
 
Signal-induced noise is observed in Hamamatsu R11265 MaPMTs. The noise is localised in specific regions of the tube periphery and extends over 3 to 4 \mus after the primary signal. It originates from combined effects of internal light emission and ion feed-back. A formalism is developed to assess the impact of this noise on the PID capabilities of the \lhcb \rich detectors during \lhc Run 3. Based on these studies, mitigation strategies are being adopted and involve the following solutions: use of an appropriate high-voltage operating point as a compromise between optimal single-photon detection efficiency and minimal number of SIN pulses; installation of new MaPMTs with reduced SIN effects in the \richone regions of highest occupancy; implementation of a nanosecond time gate in the digital readout board firmware to improve the signal-to-noise ratio. The expected \rich performance is evaluated by means of full simulation studies carried out with the \lhcb software stack. It is shown that excellent PID can be achieved after these mitigation strategies are applied.

\section*{Acknowledgements}

\noindent We would like to thank the LHCb RICH team for supporting this publication, and in particular Antonis Papanestis for the review. We express our gratitude to our colleagues in the CERN
accelerator departments for the excellent performance of the LHC. We
thank the technical and administrative staff at the LHCb
institutes.
We acknowledge support from CERN and from the national agencies:
CAPES, CNPq, FAPERJ and FINEP (Brazil); 
MOST and NSFC (China); 
CNRS/IN2P3 (France); 
BMBF, DFG and MPG (Germany); 
INFN (Italy); 
NWO (Netherlands); 
MNiSW and NCN (Poland); 
MEN/IFA (Romania); 
MSHE (Russia); 
MICINN (Spain); 
SNSF and SER (Switzerland); 
NASU (Ukraine); 
STFC (United Kingdom); 
DOE NP and NSF (USA).
We acknowledge the computing resources that are provided by CERN, IN2P3
(France), KIT and DESY (Germany), INFN (Italy), SURF (Netherlands),
PIC (Spain), GridPP (United Kingdom), RRCKI and Yandex
LLC (Russia), CSCS (Switzerland), IFIN-HH (Romania), CBPF (Brazil),
PL-GRID (Poland) and NERSC (USA).
We are indebted to the communities behind the multiple open-source
software packages on which we depend.
Individual groups or members have received support from
ARC and ARDC (Australia);
AvH Foundation (Germany);
EPLANET, Marie Sk\l{}odowska-Curie Actions and ERC (European Union);
A*MIDEX, ANR, IPhU and Labex P2IO, and R\'{e}gion Auvergne-Rh\^{o}ne-Alpes (France);
Key Research Program of Frontier Sciences of CAS, CAS PIFI, CAS CCEPP, 
Fundamental Research Funds for the Central Universities, 
and Sci. \& Tech. Program of Guangzhou (China);
RFBR, RSF and Yandex LLC (Russia);
GVA, XuntaGal and GENCAT (Spain);
the Leverhulme Trust, the Royal Society
 and UKRI (United Kingdom).

\addcontentsline{toc}{section}{References}
\setboolean{inbibliography}{true}
\bibliographystyle{LHCb}
\bibliography{main,standard,LHCb-PAPER,LHCb-CONF,LHCb-DP,LHCb-TDR}

\end{document}